\documentclass[ 
reprint,
 amsmath,amssymb,
 aps,showpacs,superscriptaddress
]{revtex4-1}

\usepackage{blindtext}
\usepackage{graphicx}% Include figure files
\usepackage{dcolumn}% Align table columns on decimal point
\usepackage{bm}% bold math
\usepackage{subfig}
\usepackage{cancel}
\usepackage{enumitem}
\usepackage{mathtools} 
\usepackage{upgreek}

\usepackage{color}
\usepackage{enumitem}

\newcommand{\beqar}{\begin{eqnarray}}
\newcommand{\eeqar}{\end{eqnarray}}
\newcommand{\bea}{\begin{eqnarray}}
\newcommand{\eea}{\end{eqnarray}}
\newcommand{\bcen}{\begin{center}}
\newcommand{\ecen}{\end{center}}
\newcommand{\ra}{\rightarrow}
\newcommand{\bra}[1]{\left< #1 \right|}
\newcommand{\ket}[1]{\left| #1 \right>}

\newcommand{\eps}{\varepsilon}
\newcommand{\lam}{\lambda}

\newcommand{\f}[2]{\frac{#1}{#2}}
\renewcommand{\b}[1]{\left({#1}\right)}
\renewcommand{\v}[1]{\vec{#1}}
\newcommand{\pd}[2]{\frac {\partial #1}{\partial #2}}

\newcommand{\brac}[1]{(#1)}
\renewcommand{\sb}[1]{\left[{#1}\right]}
\newcommand{\mean}[1]{\langle {#1} \rangle}

\begin{document}

\preprint{APS/123-QED}
\title{Inertial Theorem: Overcoming the quantum adiabatic limit}

\author{Roie Dann}
\email{roie.dann@mail.huji.ac.il}
\affiliation{The Institute of Chemistry, The Hebrew University of Jerusalem, Jerusalem 9190401, Israel}%
\affiliation{Kavli Institute for Theoretical Physics, University of California, Santa Barbara, CA 93106, USA}
\author{Ronnie Kosloff}%
\email{kosloff1948@gmail.com}
\affiliation{The Institute of Chemistry, The Hebrew University of Jerusalem, Jerusalem 9190401, Israel}%
\affiliation{Kavli Institute for Theoretical Physics, University of California, Santa Barbara, CA 93106, USA}

\begin{abstract}
We present a new theorem describing stable solutions for a driven quantum system. The theorem, coined `{\em{inertial theorem}}',  is applicable for fast driving, provided the acceleration rate is small. The theorem states that in the inertial limit eigenoperators of the propagator
remain invariant throughout the dynamics, accumulating dynamical and geometric phases.
The proof of the theorem utilizes the structure of Liouville space and a closed Lie algebra of operators. We demonstrate applications of the theorem by studying three explicit solutions of a harmonic oscillator, a two-level and three-level system models. These examples demonstrate  that the inertial solution is superior to that obtained with the adiabatic approximation. Inertial protocols can be combined to generate a new family of solutions. 
The inertial theorem is then employed to extend the validity of the Markovian Master equation to strongly driven open quantum systems. 
In addition, we explore the consequence of new geometric phases associated with the driving parameters.
\end{abstract}

%\pacs{03.65.−w,03.65.Yz,32.80.Qk,03.65.Fd}

\maketitle

\section{Introduction}
Closed-form solutions for the propagator are of utmost importance for quantum control \cite{huang1983controllability,shapiro2003principles,tannor1985control,kosloff1989wavepacket,rabitz2000whither,glaser2015training}. Generally, any Hamiltonian that allows control does not commute with itself at different times, leading to a time-ordering procedure in the evolution operator \cite{albertini2001notions,marecek2020quantum}.
Hence, the development of closed solutions, for control Hamiltonians, faces the formidable task of time-ordering \cite{dyson1949s,blanes2009magnus}. 
Moreover, one desires that such solutions are stable under variations in the driving protocol and external noise.  

%\tg{Another important requirement in quantum control is the stability of the solution, under variations in the driving protocol and external noise. }

The present paper is devoted to the construction of  closed-form, stable solutions of driven quantum systems. Currently, the popular theoretical as well as experimental approach utilizes the adiabatic theorem, constrained to slow driving \cite{schiff1968quantum,comparat2009general,mostafazadeh1997quantum,sarandy2005adiabatic,kato1950adiabatic}.
Here, we propose approximate solutions for rapid processes based on the inertial theorem. In the appropriate limit these solutions incorporate the adiabatic approximation.

%\dropcap{Q}uantum state manipulation is an integral part of contemporary quantum science \cite{cirac1995quantum,monroe1995demonstration,barreiro2011open,rosi2013fast,bloch2008many,duan2001geometric,loss1998quantum,brooke1999quantum,venegas2018cross,johnson2011quantum,rossnagel2016single,pekola2015towards}. This task is achieved by engineering the Hamiltonian by means of external driving \cite{rice1992new,glaser2015training}.
%Typically, in such processes, the Hamiltonian does not commute with itself at different times, $\sb{\hat H\b t,\hat H\b{t'}}\neq 0$. As a result, the solution of the dynamics is confronted with the obstacle of time-ordering \cite{blanes2009magnus}. 

% ($d\mu/dt \ra 0$)
The inertial theorem utilizes a timescale separation between variables to generate the system's evolution for slow acceleration of the external driving. The derivation subsides in the Liouville space and requires the existence of a time-dependent operator basis. Formally, the theorem is similar to the adiabatic theorem, where adiabatic states are replaced by the time-dependent eigenoperators of the generator of the propagator \cite{dann2018time}.% The dynamics of eigenoperaters are characterized by geometric and dynamical phases.
The inertial solutions remain precise for rapid driving of the system under the condition of slow acceleration relative to the system's dynamics. 
%Similarly to the adiabatic solution, the solutions are stable with respect to small variations in 
%the driving protocol. 

Inertial protocols can  generate a diverse family of solutions, as the fast degrees of freedom are arbitrary smooth functions while only the rate of change of the slow variables is restricted. These solutions identify new invariant operators, which are time-dependent constants of motion \cite{wulfan1981adiabatic,pfeifer1983stationary}. Moreover, inertial solutions can be combined to generate an inertial Hamiltonian, extending the family of possible solutions.

Quantum control is an integral part of contemporary quantum science \cite{cirac1995quantum,monroe1995demonstration,barreiro2011open,rosi2013fast,bloch2008many,duan2001geometric,loss1998quantum,brooke1999quantum,venegas2018cross,johnson2011quantum,rossnagel2016single,pekola2015towards}. Control is commonly achieved by engineering the Hamiltonian by means of external driving \cite{rice1992new,glaser2015training}.
%Time-dependent processes are ubiquitous in quantum science. For example, when loading and manipulating cold atoms and ions \cite{cirac1995quantum,monroe1995demonstration,barreiro2011open,bloch2008many}, generating quantum gates \cite{jaksch2000fast,jonathan2000fast,nielsen2002quantum}, quantum annealing \cite{kadowaki1998quantum,finnila1994quantum,brooke1999quantum,venegas2018cross,johnson2011quantum,santoro2002theory}, and quantum thermodynamic devices \cite{rossnagel2016single,pekola2015towards}.
 The inertial theorem  can be a crucial element of the control tool box, generating rapid control protocols which go beyond the adiabatic driving regime. 
 In addition, protocols based on time-dependent constants of motion, can replace adiabatic protocols, such as the STIRAP protocol for a three-level system \cite{oreg1984adiabatic,shore1991multilevel,vitanov2017stimulated}. 

%Ninety years ago, Born and Fock presented the quantum adiabatic theorem \cite{born1928beweis}.
%The theorem states that for a slowly varying Hamiltonian $\hat H \b t$, an eigenstate $\ket{n\b 0}$ of the initial Hamiltonian $\hat{H}\b 0$, remains an eigenstate $\ket{n \b t}$ of the instantaneous Hamiltonian $\hat{H}\b t$ throughout the process. 
%By inserting the instantaneous eigenstate solution into the time-dependent Schr\"odinger equation, the validity of the adiabatic approximation can be determined.

%The adiabatic approximation is the customary approach to
%circumvent the time-ordering and solve the quantum dynamics
%\cite{schiff1968quantum,comparat2009general,mostafazadeh1997quantum,sarandy2005adiabatic,kato1950adiabatic}. 
%The approximation is valid when the change in the Hamiltonian is small relative to the square of the  energy gaps. 
%Consequently, practical adiabatic processes require long timescales, resulting in accumulated sensitivity to environmental noise.

%The magnitude of the deviation is  quantified by the adiabatic parameter $\mu$, in the adiabatic limit $\mu\ra 0$ \cite{mostafazadeh1997quantum}.

We demonstrate applications of the inertial theorem by studying three physical models: a time-dependent harmonic oscillator, a driven two-level system, and a three-level system.
These models are the building blocks of both experimental and theoretical  studies performed in the quantum regime \cite{rossnagel2016single,cirac1995quantum}.

We utilize the inertial solution to derive the equations of a motion for a driven open quantum system. We then analyse the geometric phase associated with the inertial solution. This phase differs in its physical role from the Berry phase of the adiabatic solution. In contrast to the Berry phase, the former phase directly influences observables and can be witnessed, for non-closed circuits in the driving parameter space. 
%The geometric phase in Liouville space is demonstrated employing two-level and three-level Hamiltonians.
For  systems interacting weakly with the environment, the geometric phase induces a shift in the decay rates.

%The article is structured as follows:
%In sections \ref{sec:inertial theorem} and \ref{sec:proof} we present the inertial solution and prove the theorem. Inertial protocols are studied for explicit models in Sec. \ref{sec:Inertial examples}, and the inertial and adiabatic solutions are compared. Following is a proof of the existence of eigenoperators and the required decomposition that leads to the inertial solution, for a general system described by a compact algebra, Sec. \ref{sec:proof of existence}. In Sec. \ref{sec:open system} we extend the description to open system dynamics and analyse the geometric phase in Sec. \ref{sec:geometric phase}. We conclude in  Sec. \ref{sec:discussion} by discussing further applications and future prospects.

\section{The inertial theorem}
\label{sec:inertial theorem}

Obtaining a closed-form solution for dynamics generated by a time-dependent Hamiltonian is a difficult task. The inertial theorem constructs a solution by incorporating the time-dependence within a time-dependent operator basis and scaled time.

The derivation of the inertial theorem is conducted in Liouville space, a state space of system operators $\{\hat{X} \}$, endowed with an inner product $\b{\hat{X_i},\hat{X_j}}\equiv\text{tr}\b{\hat{X_i}^{\dagger}\hat{X_j}}$ \cite{fano1957description,von2018mathematical,scopa2019exact}. 
In Liouville space, the system's dynamics are represented in terms of a basis of orthogonal operators  $\{\hat{V}\}$, spanning the space.  For example, for a two-level-system, the operator basis can be the Pauli operators. 
An arbitrary operator $\hat{A}$, spanned by operator the basis $\{\hat{V}\}$, $\hat{A}\b t=\sum_{i=1}^{N^2} a_i\b t\hat{V}_i$, is represented by the vector $\v{A}\b t=\{a_1\b t,a_2\b t,...,a_{N^2}\b t\}^T$ in Liouville space, where $a_i\b t$ are time-dependent coefficients and $N$ is the size of the Hilbert space.  

Employing the Heisenberg equation of motion, the dynamics of systems in Liouville space are determined by
 \begin{equation}
 \f{d}{dt}\v{v}^H \b t = \hat{U}^\dagger\b{t,0}\sb{\b{i \sb{\hat{H}\b t,\bullet} +\pd{}{t}} \v{v}\b t} \hat{U}\b{t,0}~~,
 \label{dynamics Liouv}
 \end{equation}
 where $\v{v}$ the expansion elements in the basis  $\{\hat{V}\}$. The convention $\hbar=1$ is used throughout this paper, and the superscript $H$ designates that the operators $\hat{V}_i$ are in the Heisenberg picture, i.e, $\hat{V}_{i}^H\b t=\hat{U}^\dagger\b{t,0}\hat{V}_i\hat{U}\b{t,0}$, where $\hat{U}\b{t,0}$ satisfies the Schr\"odinger equation with respect to the Hamiltonian $\hat{H}\b t$ (at time $t=0$ the Schr\"odinger and Heisenberg pictures coincide, $\v v^{H}\b 0=\v{v}\b 0 $).
 
We consider a finite time-dependent basis, forming a closed Lie algebra; this guarantees that  Eq. \eqref{dynamics Liouv} can be solved within the basis \cite{alhassid1978connection}.
This property applies trivially for any finite Hilbert space,
or else, when a closed sub-algebra can be found, for example, the Heisenberg-Weyl group that defines the Gaussian states of the quantum harmonic oscillator \cite{streater1967representations}. 

It is useful to limit the description to the minimal sub-algebra required to solve the system's dynamics.
In the case of compact algebras, this greatly simplifies the analysis, while for non-compact algebras, finding a sub-algebra is a prerequisite for constructing the inertial solution.

For a closed Lie algebra, Eq. (\ref{dynamics Liouv}) has the simple form
\begin{equation}
     \f{d}{dt}\v v^H \b t = {-i {\cal{M}}\b t} \v v^H \b t~~,
     \label{Schro Liouv1}
\end{equation}     
where ${\cal{M}}$ is a $N^2$ by $N^2$ matrix with time-dependent elements and  $\v{v}^H$ is a vector \footnote{For the case of compact Lie algebras and unitary dynamics,  $\cal{M}$ is guaranteed to be Hermitian.}. For compact algebra $\cal M$ has real eigenvalues, for non-compact algebras complex eigenvalues are possible (see Sec. \ref{subsec: paremetric HO}) \cite{uzdin2013effects}.
%For compact algebras $\cal M$ is Hermitian, for non-compact algebras it can be non-Hermitian (see Sec. \ref{subsec: paremetric HO}) \cite{uzdin2013effects}. 
%Equation Eq. \eqref{Schro Liouv1} is formally analogous to the time-dependent Schr\"odinger equation for a time-dependent Hamiltonian, this can be seen by transforming  $\v v\b t \ra \ket{\psi\b t}$ and ${\cal{M}}\b t \ra \hat{H}\b t$.
%Similarly, in both cases the solution  yields the dynamics of the system.

%However, the analogy only goes this far, the two equations belong to different Hilbert spaces and act upon different mathematical entities 
%\footnote{The $\text{Schr\"odiger}$ equation belongs to the Hilbert space of wavefunctions, while Eq. \ref{Schro Liouv1} is an equation of motion of a Hilbert space of operators, the Liouville space.}. 
%Moreover, the matrix  $\cal{M}$ can be non-hermitian where typically the Hamiltonian is Hermitian. 

A formal solution for Eq. \eqref{Schro Liouv1} requires a time-ordering procedure $\v v^H \b t ={\cal{T}}\exp\b{-i\int_0^t {\cal{M}}\b{\tau}d\tau}\v{v}\b 0$, where ${\cal T}$ is the chronological time-ordering operator. This formal expression is impractical, as it includes an infinite sum of integrals \cite{dyson1949s}.

The current derivation bypasses the time-ordering procedure by the following strategy: we search for a driving protocol that allows solving Eq. \eqref{Schro Liouv1} explicitly, and then extend the solution to a broad range of protocols employing the inertial approximation. By choosing 
a suitable time-dependent operator basis, the generator of the dynamics in Liouville space can be expressed as
\begin{equation}
 {\cal{M}}\b t = {\cal P}\b{\v{\chi}} {\cal{D}}\b{\v\chi,\v{\Omega}\b t} {\cal P}^{-1}\b{\v{\chi}}~~.
 \label{eq:factorized}
\end{equation}
Here, ${\cal P}\b{\v{\chi}}$ is an  invertible matrix (unitary for an Hermitian $\cal{M}$), dependent on constant parameters $\{\chi\}$. $\cal D$ is diagonal real matrix with time-dependent eigenvalues, which are a function of both $\{\chi\}$ and time-dependent parameters $\{\Omega\b t\}$. The parameters are expressed in short notation as $\v \chi =\{\chi_1,\chi_2,...,\chi_m\}^{T}$ and $\v \Omega\b t =\{\Omega_1 \b t,\Omega_2 \b t,...,\Omega_{N^2} \b t\}^{T}$.
In the first two examples presented, there is a single parameter $\v \chi=\chi$, which is equal to the adiabatic parameter $\mu$. More general examples include multiple inertial variables, Cf. Sec. \ref{subsec:3 level}.

We will prove in Sec. \ref{sec:proof of existence} that decomposition Eq. \eqref{eq:factorized} can always be achieved for any time-dependent analytical Hamiltonian $\hat{H}\b t$. Nevertheless, the existence of a solution is not constructive, therefore, the suitable time-dependent basis, associated with a general protocol $\hat{H}\b t$ is not straightforward. Once a suitable protocol and time-dependent operator basis is found, for which $\cal M$ obtains the required form, Eq. \eqref{eq:factorized}, the solution becomes
\begin{equation}
 \f{d}{d t}\v v^H \b t  = -i {\cal P}\b{\v{\chi}} {\cal{D}}\b{\v\chi,\v{\Omega}\b t} {\cal P}^{-1}\b{\v{\chi}} \v v^H \b t~~.
\label{theta_motion} 
 \end{equation} 
 
 Next, we write the eigenvalues of ${\cal{D}}$ as a product of a $\v{\chi}$ dependent function and a time-dependent function, leading to: ${\cal{D}}= \text{diag}\b{\lam_1\b{\v \chi}\Omega_1\b t,...,\lam_N\b{\v \chi}\Omega_N\b t}$. Since $\lam_j$, $\Omega_j$ and $\v{\chi}$ are not specified, such decomposition is general and does not enforce further restrictions on our result.
The solution of Eq. (\ref{theta_motion}) is straightforward, yielding 
\begin{equation}
 \v v^H \b{t} = \sum_{k=1}^{N^2}c_k\v{F_k}\b{\v{\chi}} e^{-i \lam_k \theta_k\b t}~~,
 \label{constat v}
\end{equation}
where the scaled-time parameters are $\theta_k\b t=\int_0^t dt' \Omega_k\b{t'}$ and $c_k=\sum_{i}{\cal P}_{ik}$ are constant coefficients. The Liouville vector  $\v {F_k}$ corresponds to the eigenoperator $\hat{F}_k=\sum_{i}{\cal{P}}^{-1}_{ki} \hat{V}_i$, where ${\cal{P}}^{-1}_{ik}$ are elements of ${\cal{P}}^{-1}$. For a Hermitian $\cal M$, the eignvalues $\lambda_k$ are either zero or are pairs with equal magnitude and opposite signs.

 The structure of Eq. \eqref{theta_motion} allows an explicit solution for the dynamics, including cases where the operator basis is time-dependent. As a result, the solution circumvents the time-ordering operation. However, the approach is limited by the condition that ${\cal{P}}$ is a constant unitary operator, i.e., $\v \chi=\text{const}$. For a set basis of operators $\{\hat{V}\}$, this condition restricts the relevant driving protocols.  

For general protocols, when $\v \chi\b{t}$ varies with time, the solution can be extended to processes of slowly varying $\v \chi \b{t}$ (inertial).  
The entire proof is reported in Sec. (\ref{sec:proof}), and follows a mathematical construction similar to that of the adiabatic theorem \cite{schiff1968quantum,ibanez2011shortcuts,sarandy2005adiabatic}. 

For a state $\v{v}$, driven by a inertial protocol, the system's evolution is given by
\begin{multline}
 \v v^H\b{t}=\sum_{k=1}^{N^2} c_k\b{\v{\chi}\b t} e^{-i\int_{0 }^{t}dt'\lam_{k}\Omega_k}e^{i \phi_{k}\b t}\v{F}_{k}\b{\v \chi\b{t}}\\
  ={\cal{P}}\b{\v \chi \b t}e^{-i\int_{\theta_k\b 0}^{\theta_k \b t} {\lambda_k\b{\theta'_k}d\theta'_k}}
    {\cal{P}}^{-1}\b{\v \chi \b t}\v{v}\b 0~~,
 \label{eq:inetrial state}
 \end{multline} 
where the first exponent is determined by the dynamical phase and the second includes a new geometric phase
\begin{equation}
\phi_k\b{t}= i\int_{\v{\chi} \b{ 0}}^{\v{\chi} \b{t}} d\v{\chi}\b{\v{G}_{k},\nabla_{\v{\chi}} \v F_{k}} ~~. 
\label{eq:geometric}
\end{equation}
Here, $\v{G}_{k}$ are the bi-orthogonal partners of $\v F_{k}$.

The system's state follows the instantaneous solution determined by the instantaneous $\v{\chi} \b{t}$ and phases associated with the eigenvalues $\lam_k\Omega_k$ and eigenoperators $\hat F_k$. We restrict the analysis to the case where $\lam_k\Omega_k$ do not cross, hence, the spectrum of ${\cal{D}}$ remains non-degenerate throughout the evolution.
 Substituting the inertial solution, Eq. \eqref{eq:inetrial state}, into Eq. \eqref{theta_motion} enables assessing the  validity of the approximation. The quality of approximation is expressed in terms of the magnitude of the `inertial parameter' (Section \ref{sec:proof})
 \begin{equation}
 \Upsilon = \sum_{n,k}\Bigg|\f{\b{\v{G}_{k},\nabla_{\v{\chi}}{\cal M}\v F_{n}}}{\b{\lam_{n}\Omega_n-\lam_{k}\Omega_k}^{2}}\b{\f{d\v\chi}{dt}}^2\Bigg|~~,
 \label{eq:Upsilon}
 \end{equation}
 which implies that the inertial solution, Eq. \eqref{eq:inetrial state}, remains valid when $\v{\chi}$ follows a path in the parameter space of $\{ \chi\}$, where the eigenvalues $\lambda_k$ and $\lambda_n$ are distinct \cite{kato1950adiabatic}. 
 
The factorization of the eigenvalues of $\cal D$ recognizes two timescales. A short timescale related to the (rapidly changing) frequencies $\{\Omega\b t\}$ and a long timescale associated with the change in $\{\lambda\b{\v \chi}\}$ variables. This separation of timescales allows containing the rapid change in the protocol within the inertial solution. The rapid time-dependence of the solution is effectively absorbed into the frequencies and time-dependent operator basis. In addition, we identify the dynamical invariant operators \cite{monteoliva1994geometric,gungordu2012dynamical,levy2018noise}, which are associated to the eigenvectors $\v{F}_k$ with vanishing eigenvalues, $\lam_k=0$.

  The inertial theorem incorporates the adiabatic theorem, since for slow driving $\{\Omega\}$ are constant and $\{\chi\}$ are slowly varying, therefore the matrix ${\cal{M}}\b t$ can be diagonalized at each instant. We then obtain the eigenoperators associated with ${\cal{M}}\b t$ and the decomposition in Eq. \eqref{eq:factorized}.
%The {\em{inertial theorem}} is applicable when the dynamics of the system can be cast in the form of Eq. \eqref{theta_motion}. 
 %In the following, two such examples are presented.

\section{Inertial theorem proof}
\label{sec:proof}
The following derivation is in the spirit of the adiabatic theorem, as presented by Schiff \cite{schiff1968quantum}, and the generalization for a non-hermitian Hamiltonian has been presented by Ibanez \cite{ibanez2011shortcuts}.
We formulate the derivation in Liouville space, a Hilbert space of operators.
%with a scalar product $\b{\hat{X_i},\hat{X_j}}\equiv\text{tr}\b{\hat{X_i}^{\dagger}\hat{X_j}}$. 
These operators are defined in terms of the underlying Hilbert space of wave-functions $\hat{X} \ket{\psi}=\ket{\phi}$ with a scalar product $\left<\psi|\phi\right>$. 

For a system described by a finite algebra of operators, the Liouville generator ${\cal M}\b{\v {\chi},\v \Omega}$ is a diagonalizable rank $N^2$, parameter dependent matrix, where the elements of $\v \chi$ and $\v\Omega$ are real parameters which can be viewed as coordinates of a parameter space. 
We assume the $N^2$ instantaneous right eigenvectors of the Liouville generator $\cal{M}$ are  non-degenerate (at all times, i.e there is no level crossing). These are denoted by $\v{F}_k\b{\v\chi}$, $k=1,2,...,N^2$, and are associated with the eigenoperators of ${\cal{M}}$ which satisfy an eigenvalue equation \cite{dann2018time}
\begin{equation}
    \hat{F}_{k}^H\b t =  \hat{U}^\dagger \b{t,0} \hat{F}_{k}\b t  \hat{U} \b{t,0}=\beta_k \b t \hat{F}_{k}\b 0 ~~,
\label{eq:eigentype}
\end{equation}
where $\beta_k$ are time-dependent complex functions  \footnote{Equation \eqref{eq:eigentype} is associated with the Heisenberg equation: $\f{d}{dt}\hat{F}_{k}^H\b t=\b{i\sb{\hat{H}^H\b t,\hat{F}_{k}^H\b t}}+\b{\pd{\hat{F}_k\b t}{t}}^H=\alpha_k\b t \hat{F}_{k}^H\b t$, where $\beta_k\b t=\exp{\int_{0}^t\,dt'\alpha_k\b{t'}}$.}. For example, when decomposition \eqref{eq:factorized} holds the eigenvalues are $\beta_k\b t=\exp\b{-i\lam_k\int_0^t\,dt' \Omega_k\b t'}$. In addition, for compact algebras, the matrix ${\cal{M}}$ is Hermitian and the left and right eigenvectors are conjugates.

We introduce a second set of biorthogonal partners $\{\v{G}\b{\v \chi}\}$, satisfying 
\begin{equation}
{\cal M}\v F_{k}=\lam_{k}\Omega_k\v F_{k}\,\,\,\,\,\text{and}\,\,\,\,\,{\cal M}^{\dagger}\v{G}_{k}=\lam_{k}\Omega_k\v{G}_{k}~~.
\end{equation}
The two sets are biorthogonal, meaning that  $\b{\v{G}_{k},\v{F}_{n}}=\delta_{kn}$,
where $\b ,$ is the scalar product of the two vectors in Liouville space. 

%\tg{The quantum state is represented in Liouville space by the vector $\v v\b{\chi\b t,\theta\b t}$. The dynamics of the vector is governed by Eq. \eqref{Schro Liouv1}.}

%\tg{For a diagonalizable matrix, there exists an invertable matrix $\cal P$, such that ${\cal P}^{-1}{\cal{M}}{\cal P}$ is diagonal. This allows identifying the eigenvectors of ${\cal{M}}$ as $\v{F}_k={\cal P}^{-1}\v v$,  where the rows of ${\cal P}^{-1}$ are the left eigenvectors of ${\cal{M}}$ \cite{horn1990matrix}.}

The set of instantaneous eigenvectors constitutes a complete basis of the Liouville space, allowing to expand the quantum state in terms of the basis elements. We therefore propose a solution for Eq. \eqref{Schro Liouv1}, which is a superposition of the eigenvectors $\v{F}_{k}$ with additional dynamical and geometric phases, Eq. \eqref{eq:inetrial state}. %The solution includes two phases a dynamical and geometric phase $\phi_k\b t $, given by Eq. \eqref{eq:geometric}.
%\begin{equation}
%\v v\b{\v \chi,\v\theta}=\sum_{n}c_{n}\b{\theta}e^{{-i\int_{\theta\b 0}^{\theta\b t}{d\theta'\lam_{n}\b{\v \chi\b{\theta'}}}}}e^{i \phi_{n}\b{\theta}}\v F_{n}\b{\v \chi\b{\theta}}.
%\label{ap:eqv}
%\end{equation}
%\begin{equation}
%\phi_k= i\int_{\v{\chi} \b{0}}^{\v{\chi} \b t}\b{\v {G}_{k},\nabla_{\v{\chi}} \v F_{k}} d\v{\chi}~~. 
%\end{equation}

The orthonormal property of the eigenvectors, $\b{\v {G}_{k},\v F_{n}}=\delta_{kn}$ implies that
\begin{equation}
\b{\nabla_{\v{\chi}}\v {G}_{k},\v F_{n}}=-\b{\v{G}_{k},\nabla_{\v{\chi}}\v F_{n}}~~,
\label{eq:eq5}
\end{equation}
for all $n$ and $k$, therefore $\b{\v{G}_{k},\nabla_{\v{\chi}}\v F_{k}}$ is pure imaginary and as a result $\phi_k$ are real. Similarly, by differentiating the identity $\b{\v G_{n},{\cal M}\v F_{k}}=0$, for $n\neq k$, we obtain
\begin{equation}
\b{\v{G}_{k},\nabla_{\v{\chi}}\v F_{n}}=\f{\b{\v{G}_{k},{\nabla_{\v{\chi}}{\cal M}}\v F_{n}}}{\lam_{n}\Omega_n-\lam_{k}\Omega_k}~~.
\label{eq:eq4}
\end{equation}

We proceed by inserting Eq. \eqref{eq:inetrial state} into Eq. \eqref{Schro Liouv1}. We then project $\v {G_{k}}$ from the left, and utilize the orthogonality property and the derived identities Eqs. \ref{eq:eq5} and \ref{eq:eq4}, to obtain a set of differential equations
\begin{multline}
\f{d c_{k}}{dt}=\nabla_{\v{\chi}} c_{k} \cdot\f{d\v\chi}{dt}=\\
-\sum_{n\neq k}c_{n}\b{t}\f{\b{\v{G}_{k},{\nabla_{\v{\chi}}{\cal }}\v F_{n}}\f{d\v\chi}{dt}}{\lam_{n}\Omega_n-\lam_{k}\Omega_k}e^{-i\xi_{nk}}~~,
\label{ap:eq_ck}
\end{multline}
with $\xi_{nk} \equiv\int_{0}^{t}{dt'}\sb{\lam_{n}\Omega_n-\lam_{k}\Omega_k}-\b{\phi_n-\phi_k}$.

%Under the dynamics of Hamiltonian Eq. \eqref{eq:HamOH}, the geometric phase vanishes ({\textit{SI Appendix }}4),
%\ref{ap:eig}
%thus, only the dynamical phase contributes to the evolution.

Typically, the dynamical phase is the dominant contribution to $\xi_{nk}$, allowing to neglect the geometric phases. 
Integrating Eq. \eqref{ap:eq_ck} and solving iteratively leads to
\begin{multline}
c_{k}\b{t}\approx c_k\b{0}\\
-\sum_{n\neq k} \int_{0}^{t}d t'\,c_{n}
\f{\b{\v{G}_{k},{\nabla_{\v{\chi}'}{\cal M}}\v F_{n}} }{\lam_{n}\Omega_n-\lam_{k}\Omega_k}\f{d\v{\chi}}{dt'}\times \\ 
\exp\b{-i {\int_{\v{\chi} \b 0}^{\v{\chi} \b{t'}} d \v{\chi}'\, \b{\f{d\v{\chi}'}{d t''}}^{-1}\b{\lam_{n}\Omega_n-\lam_{k}\Omega_k}}}~~.
\label{ap:eq6}
\end{multline}
The term $\b{\f{d\v{\chi}'}{d t''}}^{-1}$ diverges in the inertial limit, inducing rapid oscillations in the last term. Assuming the integrand of the last exponent is integrable in the interval $\sb{\v{\chi} \b{0},\v{\chi}\b t}$, the Riemann-Lebesgue lemma, implies that when the phase of the last exponent, or $\xi_{nk}$, change rapidly relative to the integrand, the sum in Eq. \eqref{ap:eq6} vanishes 
\footnote{The Riemann-Lebesgue lemma can be stated as follows: Let $f:\sb{a,b}\ra\mathbb{C}$ be an integrable function on the interval $\sb{a,b}$. Then $\int_a^b dx e^{i n x} f \b x \rightarrow 0$ as $n\rightarrow \pm \infty$ }. 
This implies that the inertial  solution is valid when
\begin{equation}
\text{max}_t\Bigg|\f{\b{\v{G}_{k},\nabla_{\v{\chi}}{\cal M}\v F_{n}}}{\b{\lam_{n}\Omega_n-\lam_{k}\Omega_k}^{2}}\b{\f{d\v\chi}{dt}}^2\Bigg|\ll1~~~,
\label{eq:inertial condition}
\end{equation}
for all $n\neq k$.
%churchill1963fourier

\section{Inertial examples}
\label{sec:Inertial examples}

The first two examples, a driven harmonic oscillator and two-level system, demonstrate the theory in the simple framework of Liouville space, with a single inertial variable. These examples allow a closed-form analysis. The third example, a driven three-level system, is more involved leading to a generalization of the STIRAP process.  

\subsection{Parametric driven harmonic oscillator}
\label{subsec: paremetric HO}
To demonstrate the inertial theorem, we begin by analyzing the dynamics of a driven harmonic oscillator.
Physically, the system can be realized by  a particle in a varying harmonic potential \cite{rossnagel2016single}, and  is represented by the Hamiltonian
\begin{equation}
\hat{H}\b t = \f{\hat{p}^2}{2m} +\f{1}{2} m \omega^2\b t  \hat{q}^2~~,
\label{eq:HamOH}
\end{equation}
where $\hat{q}$ and $\hat{p}$ are the position and momentum operators, $m$ is the particle mass and $\omega\b t$ is the oscillator frequency.

 We consider an initial Gaussian state, which is fully defined by the set of time-dependent operators:
  $\hat{L}\b t=\f{\hat{p}^{2}}{2m}-\f 12\omega^{2}\b t \hat{q}^{2}$, $\hat{C}\b t=\f{\omega\b t}2\b{\hat{q}\hat{p}+\hat{p}\hat{q}}$, $\hat{K}\b t=\sqrt{\omega\b t}\hat{q}$, $\hat{J}\b t=\f{\hat{p}}{m\sqrt{\omega\b t}}$ and the Hamiltonian, Eq. \eqref{eq:HamOH} \cite{kosloff2017quantum}. This set of operators constitutes a basis of the Liouville space and fulfills the requirements for the decomposition of the generator, Eq. \eqref{theta_motion}. 
  The dynamics of a vector of ones, in the basis $\{\hat{H},\hat{L},\hat{C},\hat{K},\hat{J},\hat{I}\}^{T}$, are  given by
\begin{equation}
 \f d{dt}\v v^H\b{t}=-i\omega\b t{\cal B}\v v^H\b{t}~~,
 \label{eq:HO dynamics}
 \end{equation} 
 with
 \begin{equation}
  {\cal B}=i\sb{\begin{array}{cccccc}
\chi & -\chi & 0 & 0 & 0 & 0\\
-\chi & \chi & -2 & 0 & 0 & 0\\
0 & 2 & \chi & 0 & 0 & 0\\
0 & 0 & 0 & \f{\chi}2 & 1 & 0\\
0 & 0 & 0 & -1 & -\f{\chi}2 & 0\\
0 & 0 & 0 & 0 & 0 & 0
\end{array}}~~~,
\label{HOprop}
  \end{equation}
  where $\chi=\mu=\f{\dot{\omega}}{\omega^{2}}$.  When $\mu\ra 0$ the adiabatic solution is exact.
  
  For a constant $\mu$, $\cal{B}$ is a constant matrix, and the form of Eq. \eqref{eq:HO dynamics} is just a special case of the decomposition Eq. \eqref{eq:factorized}, where ${\cal{D}}=\text{diag}\omega\b t\b{b_1,...,b_8}$, where $b_i$ are the eigenvalues of ${\cal B}$, and $\cal{P}\b{\chi}$ is the diagonalizing matrix of $\cal{B}$. The frequencies are identified as $\Omega_k\b t=\omega\b t$ for all $k$, with a single scaled-time parameter $\theta \b t=\int_0^t\,dt'\omega\b{t'}$. 
  
 The eigenoperators and  eigenvalues,  $\hat{F}_k$ and $\lam_k$, are obtained by diagonalization (see Appendix \ref{ap:eig}).%\ref{ap:eig}}
 The matrix $\cal{B}$ has real eigenvalues that possess a non-Hermitian degeneracy for $\chi=\mu=2$ \cite{uzdin2013effects}. This limits the solution to avoid proximity to the degeneracy point.
The inertial parameter, Eq. \eqref{eq:Upsilon}, takes the form $\Upsilon \sim \b{\f{\dot{\mu}}{2\kappa\omega}}^2$, where $\kappa = \sqrt{4-\chi^2}$, which explicitly becomes
%\begin{equation}
%\Upsilon=\frac{\mu^{2}\b{\f{\ddot{\omega}}{\omega}-2\mu^{2}}}{\b{2\kappa}^{2}\left(\f{\ddot{\omega}}{\omega}\log\left(\f{\omega\b t}{\omega\b 0}\right)-\mu^{2}\left(2\log\left(\frac{\omega(t)}{\omega\b 0}\right)+1\right)\right)}.
%\label{eq:Upsilon_HO}
%\end{equation}
 \begin{equation}
    \Upsilon=\sb{\frac{1}{2\kappa\omega^{3}}\b{\ddot{\omega}-\omega^{3}\mu^{2}}}^2
    \label{eq:Upsilon_HO}~~.
     \end{equation}
 When $\Upsilon \ll1$ the inertial solution, Eq. (\ref{eq:inetrial state}), is a good approximation of the true dynamics.

For the demonstration, we consider a particle of mass $m=1$ in a varying harmonic potential. The particle is initialized in the ground state $\rho\b 0=\ket{0}\bra{0}$, associated with the initial frequency $\omega\b 0=20$.

\begin{figure}[htb!]
\centering
\includegraphics[width=0.4\textwidth]{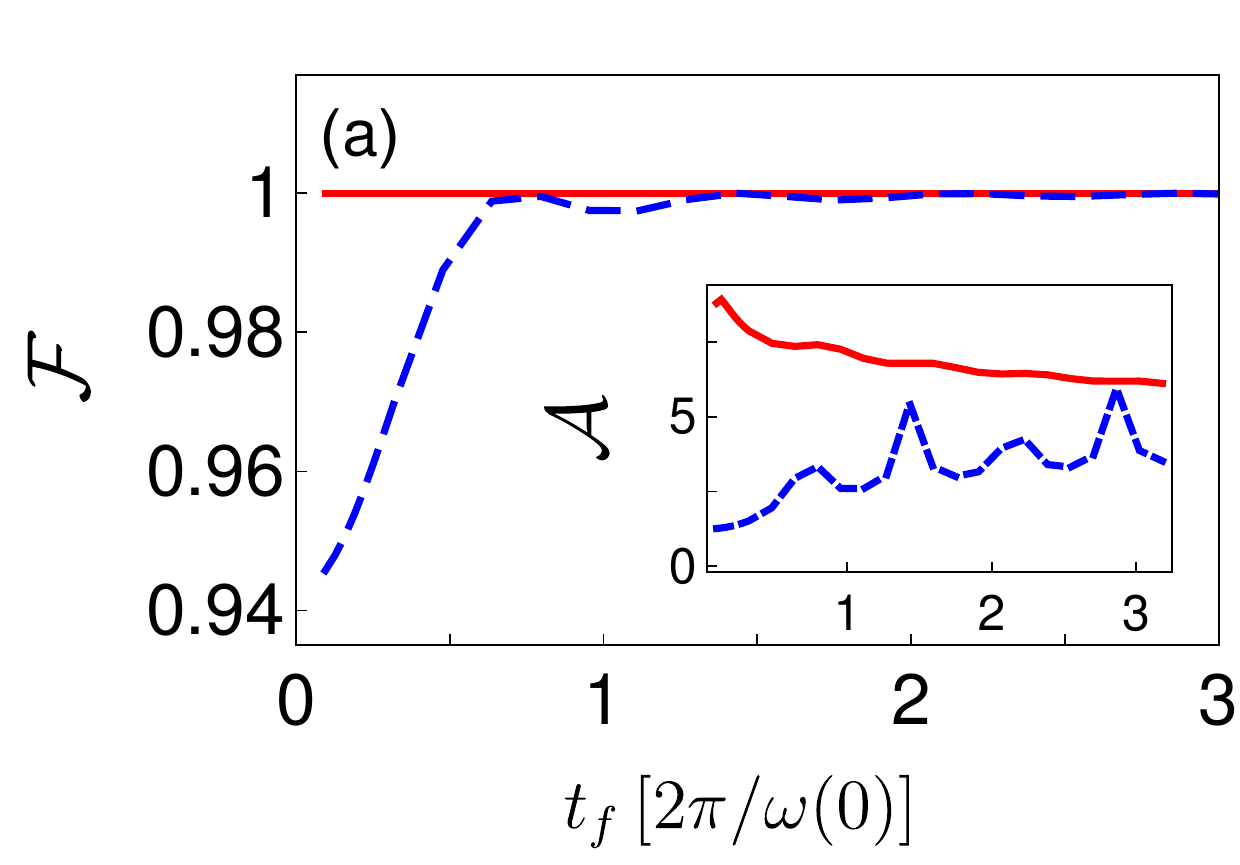}\\
\includegraphics[width=0.4\textwidth]{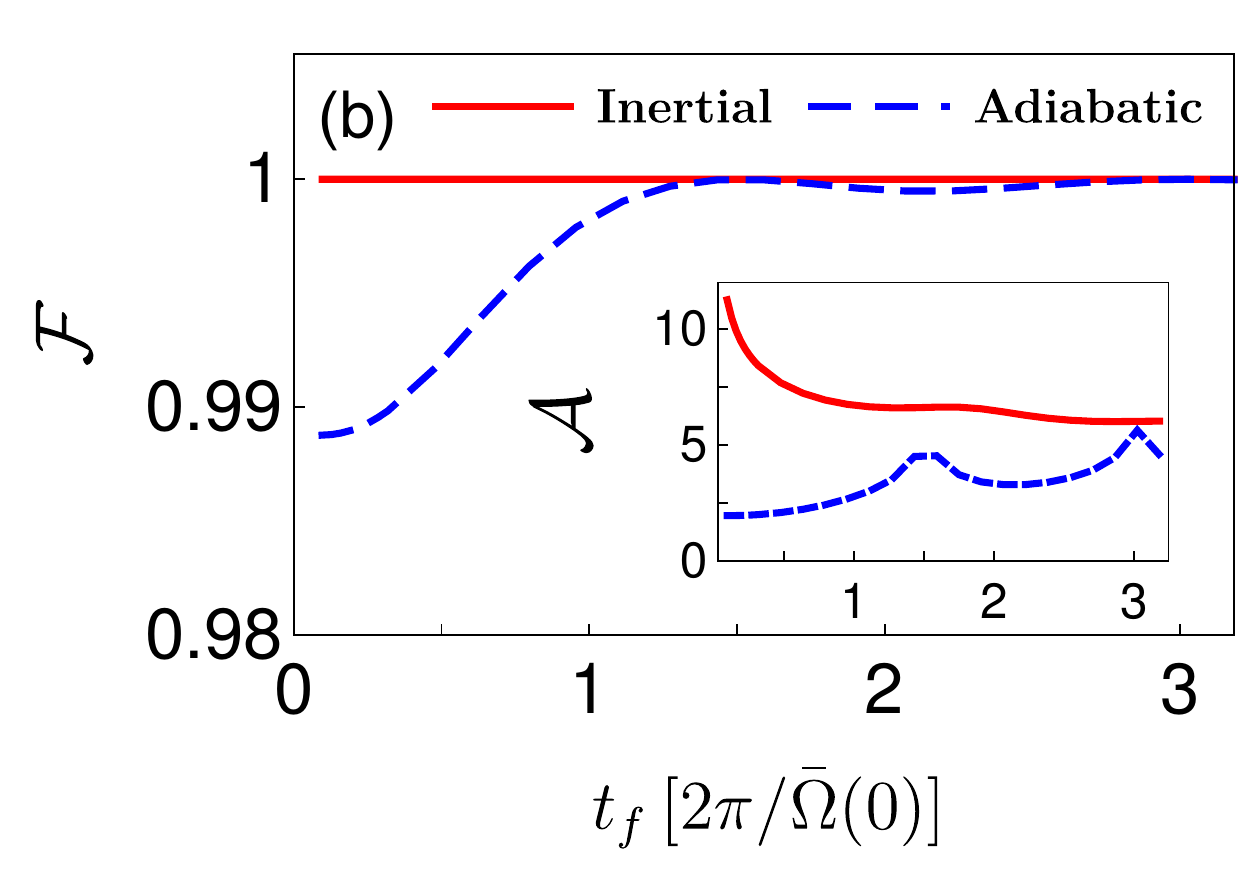}\\
\caption{The fidelity of the final state as a function of the protocol time, $t_f$, for (a) the harmonic oscillator (HO) and (b) two-level-system (TLS). As the accuracy improves, the fidelity reaches unity. (Inset) The quality measure ${\cal{A}}\equiv-\log\b{1-{\cal{F}}}$, of the inertial solution as a function of time. As the fidelity converges to unity ${\cal{A}}$ increases. The increase in the fidelity at small times can be explained by Eq. \eqref{lin chi}. As $t_f$ decreases $\chi\b{t}$ becomes constant, ($\chi\b{t_f}\ra\chi\b{0}$). Calculation parameters for the HO are:
$\omega\b 0=20$, $\omega\b{t_f}=10$, and $a=-5\cdot10^{-3}$.
The parameters for the TLS are:  $\bar{\Omega} \b 0 = 20$, $\bar{\Omega} \b{t_f}= 10$, $\eps=8$ and $\bar{a}=-5\cdot10^{-3}$, with initial state $\mean{\v v \b{0}}=\{4,1,1\}^T$.} 
\label{fig:HO} 
\end{figure}

The inertial approximation is evaluated by comparison to  a converged numerical solution, denoted by $\hat{\rho}_N$. The fidelities $\cal{F}$ of the inertial and adiabatic solutions are calculated in terms of the Bures distance with respect to $\hat{\rho}_N$, ${\cal{F}}=\sb{\text{tr}\b{{\sqrt{\sqrt{\hat{\rho}_N}\hat{\rho}\sqrt{\hat{\rho}_N}}}}}^2$  \cite{isar2009quantum}. The fidelities are compared in Fig. \ref{fig:HO}.

For the analysis, we use the protocol: $\omega\b t=\f{\omega\b 0}{1-\omega\b 0\b{\chi\b 0t+\f a2t^{2}}}$, which satisfies
\begin{equation}
    \chi\b t=\mu\b t=\chi \b 0 +a\cdot t~~.
    \label{lin chi}
\end{equation}
The protocol is designed by imposing $\omega\b{0}$, $\omega\b{t_f}$ and parameter $a$, while $t_f$ and $\chi\b 0$ are adjusted accordingly. Modifying the protocol duration $t_f$ interpolates between the sudden and adiabatic limits \footnote{Under the dynamics generated by Hamiltonian Eq. \eqref{eq:TLS hamil} the geometric phase vanishes (Appendix \ref{ap:eig})}.%\ref{ap:eig}

\begin{figure}[htb!]
\centering
\includegraphics[width=4.5cm]{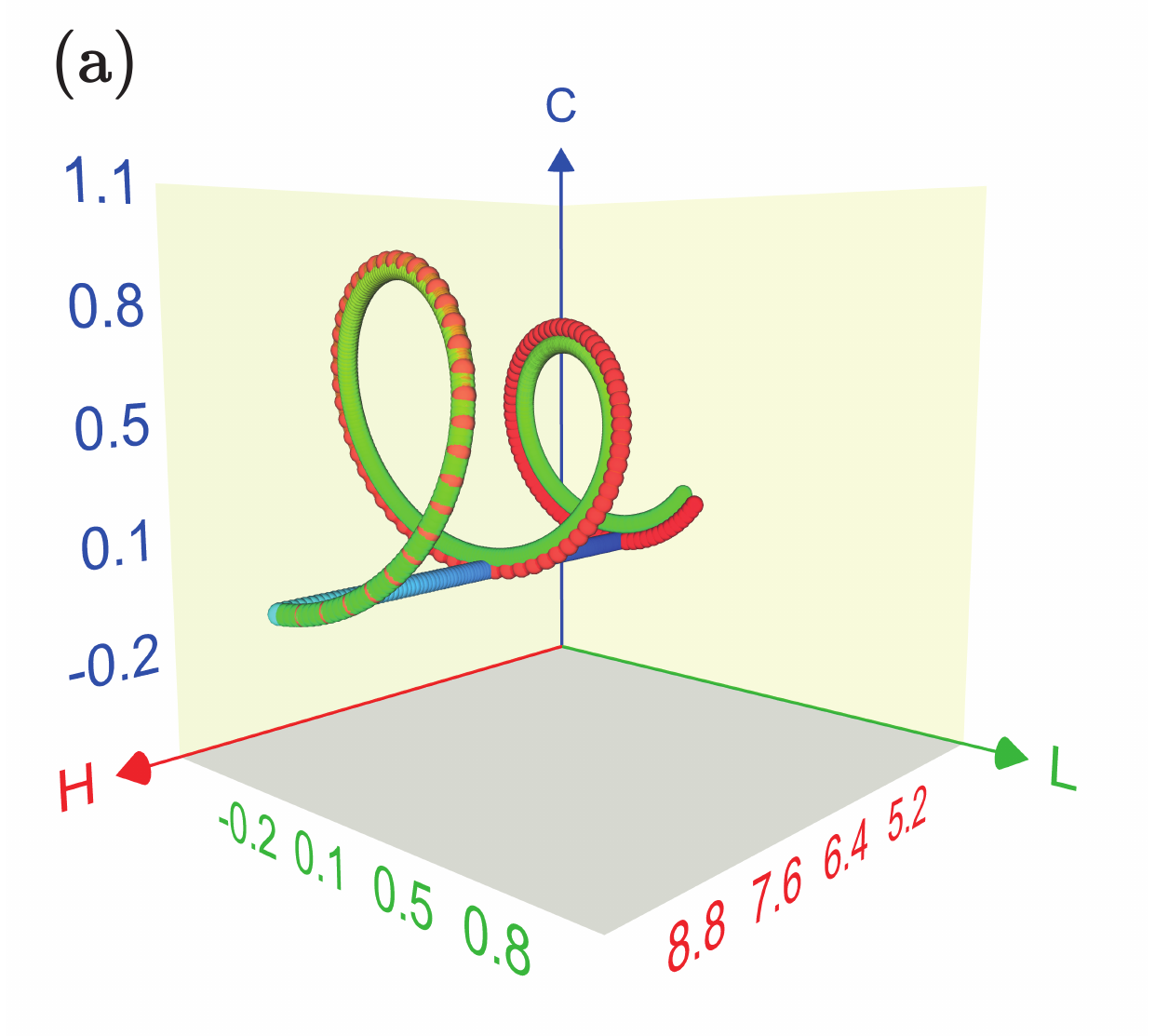}%{HO_parameter_comp.png}
\includegraphics[width=4.5cm]{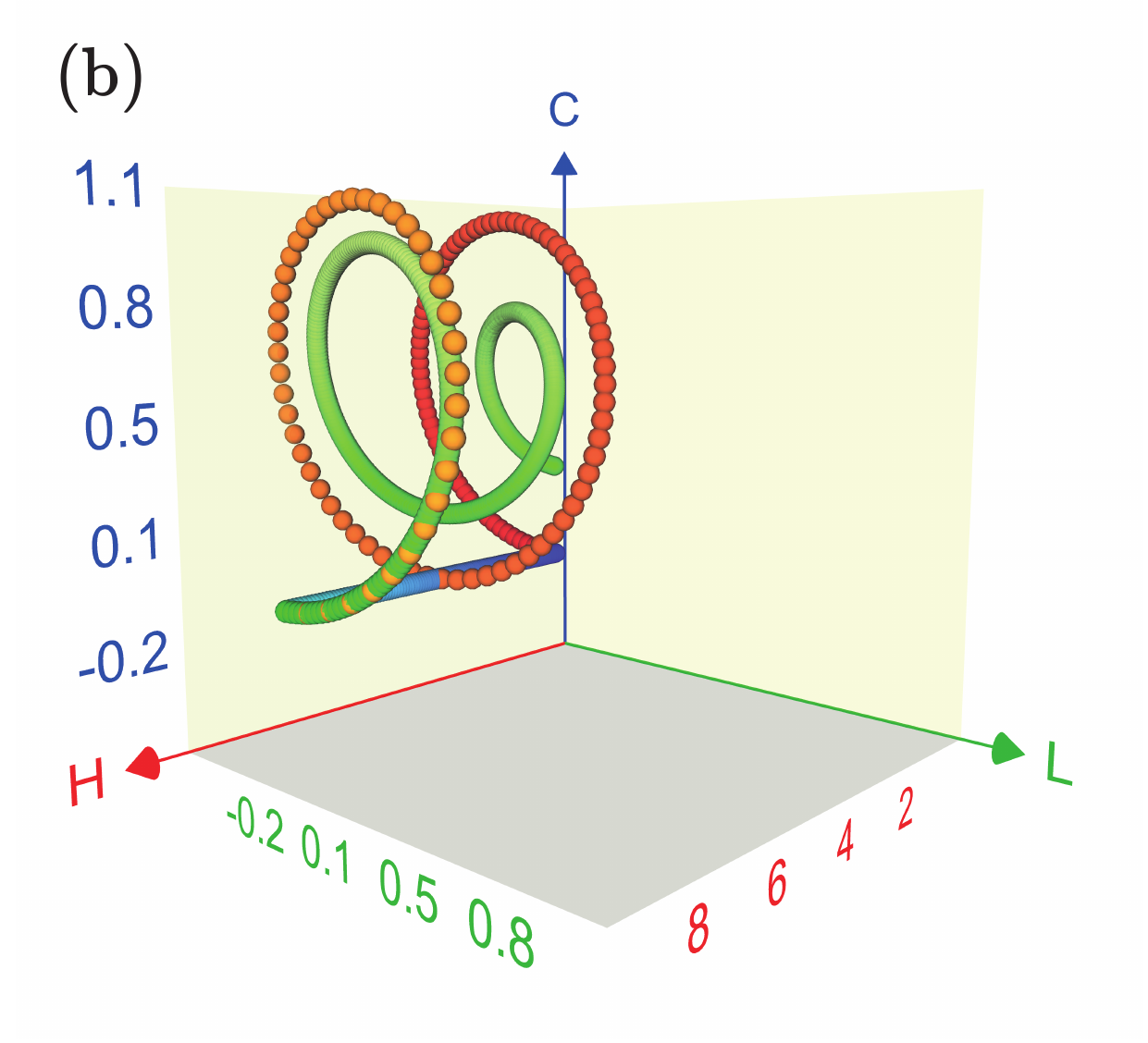}
\caption{The expectation values of $\hat{H}$, $\hat{L}$ and $\hat{C}$ (harmonic oscillator operators) for the inertial (red), adiabatic (blue) and converged numerical solution (green) for the protocol with constant `acceleration' Eq. \eqref{lin chi}. Panel (a) demonstrates the validity of the inertial solution for a protocol with small `acceleration' $\mu\b 0=-0.1$ and $a=-0.02$. Panel (b) shows the breakdown of the inertial approximation for a protocol with high `acceleration' $\mu\b 0 = -0.1$ $\alpha=-0.3$.}
\label{fig:3D1}
\end{figure}

\begin{figure}[htb!]
\centering
\includegraphics[width=7cm]{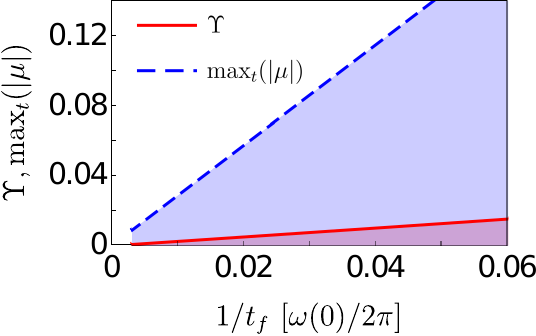}%fig_3.eps{parameters_one_over2.eps HO_parameter_comp.png}
\caption{Inertial (red) and adiabatic (blue) parameters for the harmonic oscillator as a function of $1/t_f$. We choose a protocol, Eq. \eqref{lin chi}, for which the inertial parameter is constant. $\max_t|\mu|$ is defined as the maximum adiabatic parameter for the same protocol. }. 
\label{fig:HO_par}
\end{figure}

Figure \ref{fig:HO} (a) shows the fidelity, ${\cal{F}}$, of the final state as a function of $t_f$.
The comparison indicates that the inertial approximation outperforms the adiabatic approximation. This is in accordance with Figure \ref{fig:HO_par},
demonstrating that the inertial parameter is always smaller than the adiabatic
parameter $\Upsilon<\mu$. The evolution of the state is presented in Figure \ref{fig:3D1} in terms of the expectation values $\mean{\hat{H}}$, $\mean{\hat{L}}$ and $\mean{\hat{C}}$. For small $a$ the inertial solution remains accurate relative to the converged numerical result, see Fig. \ref{fig:3D1} (a). Increasing $a$ leads to the breakdown of the inertial solution, as shown in Fig. \ref{fig:3D1} (b). The stability of the inertial approximation can be checked by adding random noise at each time-step to $\chi \b 0$ and $a$ in equation \eqref{lin chi}. As expected, the solutions were stable to Gaussian noise with a standard deviation in the scale of $\sim 0.1 \cdot a$ .

%the adiabatic parameter $\mu$ to the inertial parameter $\Upsilon$. In both cases, the validity of the approximation improves for decreasing parameter values.

%The maximum improvement factor obtains the value, $\text{max}_{t_f}\sb{\f{1-{\cal{F}}_{inert}}{1-{\cal{F}}_{adi}}}=0.66$.

%\begin{figure}[htb!]
%\centering
%\includegraphics[width=7cm]{combined4.png}
%\caption{\label{fig:HO} Measure of the accuracy of the final state as a function of the process time, $t_f$, for the harmonic oscillator (A) and two-level-system (B). %Inertial (blue) and Adiabatic solutions (red), 
%For both figures the broken line defines the protocol time where the maximal value of $\mu=0.2$, this indicates the cross-over to the adiabatic limit. As the accuracy improves the fidelity reaches unity and the measure ${\cal{A}}=-\text{log}_{10}\b{1-{\cal{F}}}$ increases.} 
%\end{figure}
% The plot of the figure is in Plot_article_2.m in HO,Inertial theorem.

%\includegraphics[width=9cm]{combined_fast.png}
%width = 0.238 for the same row

%The accuracy of the adiabatic versus the inertial solutions } quantified by comparing 

A general protocol $\omega \b t$ with a duration $\tau$ can be expressed in terms of a dimension-less parameter $s=t/\tau$. The adiabatic parameter then becomes $\mu\b s=\b{\omega\b s -\omega\b 0}/\b{\omega\b s\omega\b{0} s \tau}$. For a sufficiently large protocol duration the inertial condition is satisfied and the inertial solution constitutes an accurate solution for the system dynamics.

The relative accuracy of the associated inertial and adiabatic solutions can be compared using $\Upsilon$, 
Eq. \eqref{eq:Upsilon_HO} and  $\mu$.  When the oscillator frequency is accelerated slowly ($\ddot{\omega}<\omega^{3}\mu^{2}$, Cf. Eq. \eqref{eq:Upsilon_HO})  the inertial parameter behaves as  $\Upsilon \propto \mu^2$. Hence, for slow acceleration, even in the adiabatic regime the inertial solution possesses superior accuracy.

%\tb{In the inertial limit $\dot{\mu}\ra 0$, leading to $\Upsilon \propto \mu^2 \dot{\mu}$. Therefore, even in the adiabatic limit the inertial solution is more accurate.}  

%For a linear protocol (or when $\ddot{\omega}\ll 1$) the inertial parameter obtains the simple form $\Upsilon={\Lambda}|\mu|$, where $\Lambda\equiv |\mu|\b{2 \kappa}^{-2}\left(2\log\left(\frac{\omega(t)}{\omega\b 0}\right)+1\right)^{-1}$. Hence, for increasing linear protocols the inertial approximation is insured to outperform the adiabatic one. 

\subsection{Two-Level-System model}
\label{sec:TLS1}
The driven two-level system, characterized by the SU(2) algebra, is utilized as  an additional demonstration of the inertial theorem. The system Hamiltonian reads
\begin{equation}
    \bar{H}\b t=\omega\b t\hat{S}_{z}+\eps\b t\hat{S}_{x}~~,
    \label{eq:TLS hamil}
\end{equation}
 where $\hat{S}_i$ is the $i=x,y,z$ spin operator, and the time-dependent Rabi frequency reads $\bar{\Omega}\b t=\sqrt{\omega^{2}\b t+\eps^{2}\b t}$.
 
 The dynamics of the system is analyzed employing a time-dependent operator basis $\{\bar{H},\bar{L},\bar{C},\hat{I}\}$, with $\bar{H}\b t=\omega\b t\hat{S}_{z}+\eps\b t\hat{S}_{x}$, $\bar{L}\b t=\eps\b t\hat{S}_{z}-\omega\b t\hat{S}_{x}$, $\bar{C}\b t=\bar{\Omega} \b t \hat{S}_{y}$.
The equation of motion for the Liouville vector for ones in the basis $\{ \bar{H},\bar{L},\bar{C}\}^{T}$ is of the form
\begin{equation}
\f 1{\bar{\Omega}}\f d{dt}\v r^H=\f{\dot{\bar{\Omega}}}{\bar{\Omega}^{2}}\bar{\cal I}\v r^H-i\bar{{\cal{B}}}\v r^H~~,
\label{eq motion TLS 1}
 \end{equation}
where
\begin{equation}
\bar{{\cal{B}}}=i
\sb{\begin{array}{ccc}
0 & \bar{\chi} & 0\\
-\bar{\chi} & 0 & 1\\
0 & -1 & 0
\end{array}}~~.
\label{TLS propagator}
\end{equation}
Here, $\v{\chi}=\bar{\chi}=\bar{\mu}=\f{\dot{\omega}\eps-\omega\dot{\eps}}{\bar{\Omega}^{3}}$, where $\bar{\mu}$ is the adiabatic parameter of the two-level-system. 
To transform Eq. \eqref{eq motion TLS 1} to the factorized form, Eq. \eqref{eq:factorized}, we introduce a scaled vector
\begin{equation}
\v u^H\b{t}\equiv \v r^H\b{t} e^{-\int_{0}^{t}\, dt' \b{\dot{\bar{\Omega}}/ \bar{\Omega}}}~~,
\end{equation}
for which the dynamics obtains the desired form, $\f{d}{d t}\v u^H = -i \bar{\Omega} \bar{{\cal B}} \v u^H$. This procedure is not limited to the two-level-system and relies on the fact that the identity $\cal{I}$ commutes with any operator.
The inertial solution is obtained by diagonalizing $\bar{{\cal{B}}}$, Eq. \eqref{TLS propagator}, leading to the form of Eq. \eqref{eq:inetrial state} (Appendix \ref{ap:eig}).%\ref{ap:eig}

We consider a protocol with a constant $\eps$ and a linear change in $\bar{\chi}$, $\bar{\chi}\b t=\bar{\mu}\b t=\bar{\chi}\b 0 + \bar{a}\cdot t$. This leads to the following protocol $\omega\b t=\eps\f{z\b t}{\sqrt{1-z^{2}\b t}}$, where $z\b t=\eps\sb{\chi\b 0 t+\f{\bar{a}}{2}\cdot t^{2}+\f{\omega\b 0}{\eps\bar{\Omega}\b 0}}$. Using this protocol, the exact adiabatic and inertial solutions were calculated. The results are shown in Fig. \ref{fig:HO}b, illustrating the superiority of the inertial solution over the adiabatic result.

An experimental verification of the inertial solution of a two-level system has been demonstrated experimentally, employing an Ytterbium ion $^{171}$Yb$^+$ in a Paul trap \cite{hu2019experimental}. The inertial protocol was realized in the experiment, demonstrating high accuracy of the inertial solution with respect to the measurements. In addition, deviations from the inertial solution were explored. Showing that when the studied protocol slightly deviates from the inertial condition, the phase still follows the inertial solution. This phenomena is witnessed as well in Fig. \ref{fig:3D1} for the harmonic oscillator. The experiment confirms the stability of the inertial solution under external noise.

\subsection{Three-level system model}
\label{subsec:3 level}

The three-level-atom is one of the most extensively studied driven systems
\cite{hioe1982nonlinear,hioe1981n,de1963octet}.
This is an elementary example of an SU(3) algebra, which is abundant in many branches of physics \cite{gell1962caltech,ne1961derivation,pontecorvo1968neutrino,nakagawa1987geometrical,naumov1994berry}. In addition, the model serves as a template for a basic experimental technique in atomic and molecular physics:  Stimulated Raman Adiabatic Passage (STIRAP). Based on an adiabatic approach and a dynamical symmetry, the STIRAP is a technique to efficiently transfer population between two quantum states via an intermediate state \cite{oreg1984adiabatic,shore1991multilevel,vitanov2017stimulated}.
Amendments to the adiabatic scheme have been suggested, forcing the system to follow the adiabatic evolution, by adding counter-adiabatic terms to the Hamiltonian. \cite{baksic2016speeding,zhou2017accelerated}.
New insight on this well established system can be gained from the inertial approach based on new invariants.

\begin{figure}[htb!]
\centering
\includegraphics[width=8cm]{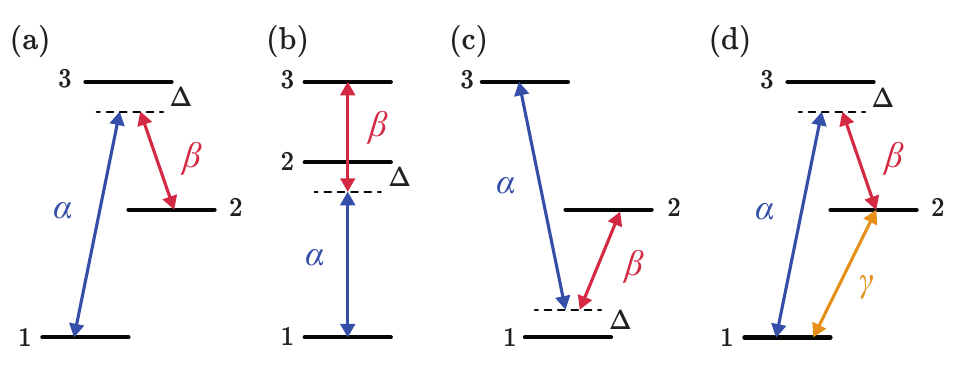}
\caption{Energy level diagram for a $\Lambda$ system, Panel (a), ladder linkage, Panel (b) and  "V" system, Panel (c), corresponding to the Hamiltonian in Eq. \eqref{eq:3level_Ham}. Panel (d) describes a fully connected Lambda system, associated with the Hamiltonian in Eq. \eqref{eq:3level_Ham2}.} 
\label{fig:3LS diagram}
\end{figure}

The basic model considers an atom with allowed transitions between the first and second levels, as well as the second to third levels.  The atom interacts with an incident electromagnetic field, coupling the levels ($\ket{1}\leftrightarrow\ket{2}$, $\ket{2}\leftrightarrow\ket{3}$): $\v E\b{z,t}=\v{\cal E}_{12}e^{i\b{\nu_{12}t-k_{12}z}}+\v{\cal E}_{23}e^{i\b{\nu_{23}t-k_{23}z}}+\text{c.c}$\,. Under the two-photon resonance condition (Gell-Mann symmetry for $N=3$) \cite{gell1962caltech,ne1961derivation,hioe1982nonlinear,hioe1981n}, the Hamiltonian within the rotating frame approximation has the form:
\begin{equation}
    \hat{H}\b t=-\sb{\begin{array}{ccc}
0 & \alpha \b t & 0\\
\alpha \b t& \Delta\b t & \beta\b t\\
0 & \beta\b t& 0
\end{array}}~~,
    \label{eq:3level_Ham}
\end{equation}
with $\Delta =\Delta_{21}=\Delta_{32}$, where $\Delta_{ij}=\nu_{ij}-\omega_{ij}$ is the detuning between the laser frequency $\nu_{ij}$ and the Bohr frequency $\omega_{ij}$. In the adiabatic STIRAP, $\Delta$ remains constant throughout the procedure.  
The Rabi frequencies are defined as $\alpha\b t = {\v d_{12}\cdot {\cal E}_{12}\b t}$ and $\beta\b t = {\v d_{23}\cdot {\cal E}_{23}\b t}$, where $\v{d}_{ij}$ is dipole moment between levels $i$ and $j$. The two-photon resonance model describes either a $\Lambda$ or ladder linkage pattern, where the photon energies correspond to the energy gap between state $\ket{3}$ and $\ket{1}$, that is,
$\omega_{31}=\nu_{21}-\nu_{23}$ for the $\Lambda$ system and $\omega_{31}=\nu_{21}+\nu_{23}$ for the ladder linkage, Fig. \ref{fig:3LS diagram}, Panels (a)-(c).

With the aim of constructing an inertial solution for Hamiltonian Eq. \eqref{eq:3level_Ham}, we begin by analyzing this model in Liouville space.
The SU(3) algebra is characterized by eight orthogonal operators along with the identity. As a consequence, the solution requires a methodical approach to obtain a suitable protocol and the eight time-dependent operators of the Liouville basis which satisfy the decomposition Eq. \eqref{eq:factorized}. For this end, we introduce a static operator basis for the SU(3) algebra, composed of trace-less Gell-Mann operators $\{\hat{\lam}_1,...,\hat{\lam}_8\}$ and the identity \cite{mann1964eightfold}. 
As required, the time-dependent Hamiltonian is within the algebra, and can be expressed in terms of the operator basis: $\hat{H}\b t=\alpha \b t\hat{\lam}_1+\beta\b t \hat{\lam}_6-\b{\Delta\b t/2}\hat{\lam}_3+\b{\Delta\b t/2\sqrt{3}}\hat{\lam}_8+\b{\Delta\b t/3}\hat{I}$.

In the next step, we search for a suitable unitary time-dependent transformation that defines the time-dependent operator basis: $\hat{T}_i\b t=\hat{S}^\dagger\b t \hat{\lam}_i\hat{S}\b t$.  Choosing the transformation
\begin{equation}
    \hat{S} = \f{1}{\Omega}\sb{\begin{array}{ccc}
\alpha & 0 & \beta\\
0 & \Omega & 0\\
\beta & 0 & -\alpha
\end{array}}~~,
\label{eq:strans}
\end{equation}
where $\Omega=\sqrt{\alpha^2+\beta^2}$, leads to the desired decomposition of the equation of motion, Eq. \eqref{eq:factorized} \cite{hioe1985gell}. This transformation conserves the simple commutation relations between the Gell-Mann matrices, while rotating the basis operators with the Hamiltonian. 
A similar choice was chosen by Eberly and Hioe \cite{hioe1985gell}, with the restriction $\alpha\b t\propto \beta \b t$.

The dynamics of the vector $\v v^H\b t$ in the $\{\{\hat{T}\},\hat{I}\}$ basis in Liouville space is generated by $\f{d}{dt}\v v^H= -i{\cal{M}}\b t\v v^H$, where ${\cal M}\b{\chi_1\b t,\chi_2\b t}$ is given in Appendix \ref{ap:eig} %\ref{ap:eig}
with $\chi_1=\Delta/\Omega$ and $\chi_2=\b{\alpha\dot{\beta}-\dot{\alpha}\beta}/{\Omega^3}$ is the adiabatic parameter.
 
%\begin{equation}
%\tb{\cal M}=
%\label{eq:3level_M}
%\end{equation}
%\begin{equation}
%{\cal M}= i\Omega\sb{\begin{array}{cccccccc}
%0 & -\chi_1 & 0 & 0 & 0 & -\chi_2 & 0 & 0\\
%\chi_1 & 0 & 2 & 0 & 0 & 0 & \chi_2 & 0\\
%0 & -2 & 0 & -\chi_2 & 0 & 0 & 0 & 0\\
%0 & 0 & \chi_2 & 0 & 0 & 0 & 1 & \sqrt{3}\chi_2\\
%0 & 0 & 0 & 0 & 0 & -1 & 0 & 0\\
%\chi_2 & 0 & 0 & 0 & 1 & 0 & \chi_1 & 0\\
%0 & -\chi_2 & 0 & -1 & 0 & -\chi_1 & 0 & 0\\
%0 & 0 & 0 & -\sqrt{3}\chi_2 & 0 & 0 & 0 & 0
%\end{array}}~~,
%\label{eq:3level_M}
%\end{equation}

When $\chi_1$ and $\chi_2$ are constant, an exact solution is obtained by diagonalization. For slowly varying $\chi_1$ and $\chi_2$ the evolution is approximated by the inertial solution Eq. \eqref{eq:inetrial state}, where  $\lam_k$ and $\v{F}_k$ are eigenvalues and eigenvectors of $\cal{M}$, and $\Omega_k=\Omega$ for all $k$. %The obtained inertial solution has two `slow' parameters, that is, $\v \chi=\{\chi_1,\chi_2\}^T$.

\subsection{Adiabatic and inertial STIRAP in Liouville space}
\label{subsec:STIRAP}

 In the three-level basis, the STIRAP procedure is a technique to completely transfer population between the states $\ket{1}$ and $\ket{3}$. This procedure is immune to losses from spontaneous emission originating from the intermediate state, $\ket{2}$, and is robust under small variations of the experimental parameters \cite{vitanov2017stimulated}. It relies on the two-photon resonance condition, $\Delta =\Delta_{21}=\Delta_{32}$, and adiabatic dynamics. This is achieved when the adiabatic parameter is sufficiently small $\chi_2\ll1$.

A dynamical symmetry viewpoint serves as an elegant approach to understand the STIRAP procedure. Within this framework, the STIRAP is a consequence of the dynamical invariance of $\hat{T}_8\b t$, when the dynamics are sufficiently slow. In the adiabatic limit, ($\chi_2\ra0$) implying that $\f{d}{dt}\hat{T}_8^H\ra0$, which means that its expectation value is constant \footnote{Equation \eqref{eq:T8 adiabatic} is obtained by eliminating $\rho_{22}$, using another constant term $\rho_{11}+\rho_{22}+\rho_{33}=1$.}
\begin{multline}
    \mean{\hat{T}_8}=-\f{\sqrt{3}}{{\Omega^{2}}}\b{{\beta^{2}}\rho_{11}+{\alpha^{2}}\rho_{33}}\\
    +\frac{\sqrt{3}\alpha\beta}{\Omega^{2}}\b{\rho_{31}+\rho_{13}}+\f 1{\sqrt{3}}=\text{const}~~.
    \label{eq:T8 adiabatic}
\end{multline}
Here, $\rho_{ij}=\bra{i}\hat{\rho}\ket{j}$, $\rho_{ii}$ is the population of the $i$'th level and $\rho_{ij}$ are the coherences ($i\neq j$). 
Any initial state $\hat{\rho}\b 0$, which is a linear combination of $\hat{I}$ and $\hat{T}_8\b 0$ is therefore also a dynamical invariant. This form incorporates the system state during the STIRAP procedure
\begin{multline}
  \hat{\rho}\b t=\f 13\hat{I}-\f 1{\sqrt{3}}\hat{T}_{8}\b t\\=  \f 1{\Omega^{2}\b t}\sb{\begin{array}{ccc}
\beta^{2}\b t & 0 & -\alpha\b t\beta\b t\\
0 & 0 & 0\\
-\alpha\b t\beta\b t & 0 & \alpha^{2}\b t
\end{array}}~~,
\label{eq:rho_adi}
\end{multline}
which maintain this form throughout the dynamics.
In such a process, the Rabi frequencies $\alpha\b t$ and $\beta\b t$ determine the boundary conditions.

The population transfer, in the adiabatic regime, is obtained by the following protocol: At initial time, $\alpha=0$ and $\beta>0$, implying that only the first state is populated, $\hat{\rho}\b 0=\ket{1}\bra{1}$, Eq. \eqref{eq:rho_adi}. This is manifested by $\mean{\hat{T}\b{\alpha=0,\beta>0}}\propto \rho_{11}$ as shown in Eq. \eqref{eq:T8 adiabatic}. During intermediate times $\alpha,\beta>0$ leading to generation of coherences between states $\ket 1$ and $\ket 3$ and rise in population of the third level. This can be witnessed by the non-vanishing terms, proportionate to $\rho_{13}$ and $\rho_{31}$ and $\rho_{33}$, in $\mean{\hat{T}_8}$, and non-vanishing off-diagonal terms in $\hat{\rho}\b t$. At the final time $t=T$, $\alpha>0$ and $\beta=0$, completing a transition of the system towards $\hat{\rho}\b T=\ket{3}\bra{3}$. The form of adiabatic invariant, Eq. \eqref{eq:rho_adi}, rationalizes the counter intuitive order of pulses of the STIRAP driving protocol \cite{vitanov2017stimulated} 

The STIRAP technique, described above, is based on the conservation of $\hat{T}_8\b t$, which applies only in the adiabatic limit ($\chi_2\rightarrow0$). In the following, we show that the STIRAP can be generalized to the inertial
regime (only requiring: $\dot{\chi_1},\dot{\chi_2}\rightarrow 0$). This technique is based on general dynamical invariants, which incorporate the adiabatic invariants as a special case.

Diagonlizing $\cal M$ leads to two independent dynamical invariants  along with the identity operator \footnote{${\hat{C}\b t}$ and ${\hat{D}\b t}$ are the two eigenoperators associated with the eigenvectors of the generator $\cal M$ with eigenvalues zero.}
\begin{equation}
 \hat{C} = \frac{\sqrt{3}\chi_2^{2}}{\chi_1}\hat{T}_{1}+\frac{\sqrt{3}\chi_2\left(\chi_1^{2}-\chi_2^{2}\right)}{\chi_1}\hat{T}_{5}-\sqrt{3}\chi_2 \hat{T}_{7}+\hat{T}_{8}~~,
\end{equation}
and $\hat{D}$, given in Appendix \ref{ap:eig}. For slowly varying $\chi_1$ and $\chi_2$ the inertial solution applies, implying that $\mean{\hat{C}\b t}$ and $\mean{\hat{D}\b t}$ are constant. 
%In addition, any linear combination of $\hat{C}\b t$, $\hat{D}\b t$ and $\hat{I}$ is also invariant. 

The operator $\hat{C}$ serves as a generalization of $\hat{T}_8$ and converges to it in the adiabatic limit ($\chi_2\ra 0$) $\hat{C}$. We can employ this property to construct an inertial STIRAP, by setting the same boundary condition as the adiabatic process, while  uplifting the restriction of slow driving at intermediate times. In this procedure, the adiabatic parameter $\chi_2$ can be large as long as the change in $\chi_2$ is sufficiently slow.

The inertial STIRAP is achieved by the following protocol: At initial time, the Rabi frequencies are set as $\alpha\b 0=0$, $\beta\b 0>0$, $\chi_2\b 0=0$ and $\chi_1\neq 0$, and the system is initialized in the $\ket{1}$ state. 
Under inertial driving, these initial conditions imply that the system remains in the form
\begin{equation}
     \hat{\rho}\b t=\f 13\hat{I}-\f 1{\sqrt{3}}\hat{C}\b t~~,
\end{equation}
throughout the dynamics. 
From an initial stationary state ($\chi_2=0$), the driving is accelerated, leading to $\alpha,\beta>0$ and $\chi_2\sim1$ at transient times, and decelerated at the final stage, achieving $\alpha>0$, $\beta=0$ and $\chi_2=0$ at the final time. Such a protocol transfers population between states $\ket{1}$ and $\ket{3}$.

The inertial STIRAP is expected to share the same robustness as the adiabatic STIRAP.
As a simple demonstration we consider a delta-correlated noise in timing of the driving. Such a process is equivalent to adding random noise to the generalized Rabi frequency $\Omega\b t$, Eq. \eqref{eq:strans} \cite{feldmann2006quantum,kosloff2010optimal}. The effective equation of motion become 
\begin{equation}
         \f{d}{dt}\v v^H \b t = -\sb{{i {\cal{M}}\b t} + \Gamma_n^2 {\cal{M}}^2\b t}\v v^H \b t~~,
     \label{noise E of M}
\end{equation}
where $\Gamma_n$ is proportional to the noise amplitude. In this case, the noise has no effect on the eigenoperators with vanishing eigenvalues (the time-dependent constants of motion \footnote{
The Heisenberg equation of motion of the `time-dependent constants of motion' operators vanishes by definition.}). The dynamics of the transient eigenoperators $\hat{F}_k$ are accompanied by an additional decay at rate $\Gamma_n^2\lam_k^2$, while the phase remains unaffected. 

The SU(3) framework can be employed to generalize this scheme, by constructing inertial STIRAP protocols for a $N$-level Hamiltonian. This can be achieved by utilizing the same techniques used to generalize the adiabatic STIRAP \cite{vitanov2017stimulated,malinovsky1997simple}.

\subsection{A fully connected three-level system}
\begin{figure}[htb!]
\centering
\includegraphics[width=7cm]{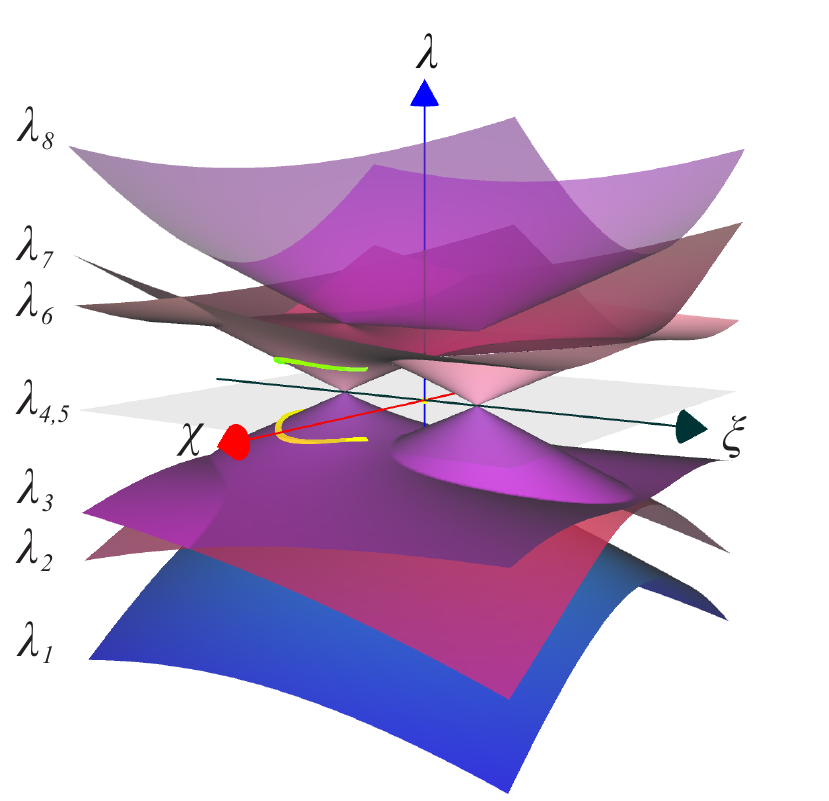}
\caption{Landscape of the eigenvalues  $\{\lam\}$ of $\cal M$ of the Hamiltonian in Eq. \eqref{eq:3level_Ham2} as a function of $\chi$ and $\xi$ for $\gamma=1/2$. The yellow and green streaks correspond to the studied inertial protocol. The protocol parameters are: $\Omega=1.5$, $\gamma=\Delta=0.5$, $\chi\b t=\sin\b{a t+\pi/4}$, $\xi=\cos\b{a t +\pi/4}$ with $a=0.01$. The driving amplitudes are $\alpha\b t=\Omega \sin\b{z\b t}$, $\beta\b t=\Omega \cos\b{z\b t}$, where $z\b t=-\Omega\b{\sin{a t}+\cos\b{a t}}/\sqrt{2} a$. }
\label{fig:geom_surface1}
\end{figure}

A similar framework of the SU(3) algebra is employed to describe the dynamics of a fully connected three-level system Fig. \ref{fig:3D1}. We study a system with the time-dependent Hamiltonian
\begin{equation}
    \hat{H}\b t =\f 1{\Omega^{2}}\left[\begin{array}{ccc}
\beta^{2}\delta & -\alpha\Omega^{2} & -\alpha\beta\delta\\
-\alpha\Omega^{2} & -\Delta\Omega^{2} & -\beta\Omega^{2}\\
-\alpha\beta\delta & -\beta \Omega^{2} & \alpha^{2}\delta
\end{array}\right]~~,
    \label{eq:3level_Ham2}
\end{equation}
where all the Hamiltonian parameters are time-dependent.
We use the transformation generated by  Eq.(\ref{eq:strans}) to define the dynamical operator basis. The dynamics in Liouville space are generated by the matrix $\cal{M}$,  similar to the propagator in Sec. \ref{subsec:3 level}, including additional inertial variables.
These are: $\chi=\b{\alpha\dot{\beta}-\dot{\alpha}\beta}/\Omega^3$, $\gamma=\Delta/\Omega$ and $\xi=\delta/\Omega$.  The landscape of the eigenvalues $\{\lam\}$ of the $\cal M$ matrix is presented in Fig. \ref{fig:geom_surface1}. For a compact algebra $\cal{M}$ is hermitian, in this case, the eigenvalues can be classified to a set of $N=3$ invariants ($\lam_i=0$) and $N\b{N-1}/2=3$ pairs of eigenvalues of the same magnitude and opposite sign. 
We find a multitude of conical intersections; for example in Fig. \ref{fig:geom_surface1} there is a series of conical intersections when $\gamma=0.5$, $\chi=0$ and $\xi=0.78,\,-1.23$.  

We consider an inertial protocol varying $\chi$ and $\xi$ slowly, with an initial state including a superposition of two eigenvectors in Liouville space: $\hat{\rho}\b 0=\f{1}{3} \hat{I}+\hat{v}_3 \b 0+\hat{v}_6 \b 0$ ($\hat{v}_i$ is the eigenoperator of $\lam_i$). The trajectory is shown 
in Fig. \ref{fig:geom_surface1} superimposed on the
eigenvlaue surfaces.

\begin{figure}[htb!]
\centering
\includegraphics[width=7cm]{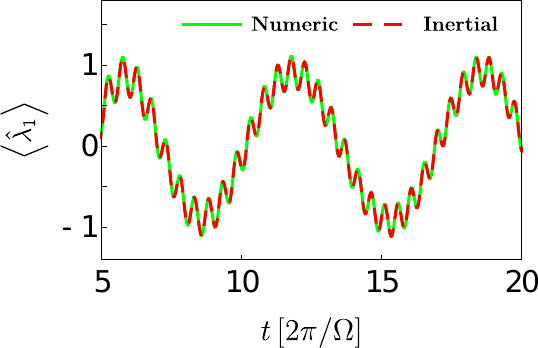}%non_closed_lam_1.pdf
\caption{ Expectation valued of the first Gell-Mann operator $\hat{\lam}_1$ ($\sigma_x$ of 1 and 2 states) as a function of time, for a system represented by the Hamiltonian in Eq. \eqref{eq:3level_Ham2}. The presented result is a typical behaviour of any system expectation values under the inertial protocol. The deviation between the numerical and inertial solutions is less than 0.015. The model parameters are given in the caption of Fig. \ref{fig:geom_surface1}.}
\label{fig:non_closed_compare}
\end{figure}

Figure \ref{fig:non_closed_compare} compares the inertial solution to a numerical integration of the equations of motion. For Hamiltonian Eq. \eqref{eq:3level_Ham2} the geometrical phase in Liouville space is identically zero since the Berry connection vanishes, Cf. Appendix \ref{ap:eig}. % \ref{ap:eig}

The demonstrated inertial protocol can be utilized in quantum control, extending the adiabatic protocols studied in Ref. \cite{augier2019semi}, for a landscape of conical intersections.

%\begin{figure}[htb!]
%\centering
%\includegraphics[width=7cm]{Compare_TLS.jpg}
%\caption{ \trr{maybe unit with the HO figure}, \tg{Measure of the accuracy of the final state of a two-level-system as a function of the process time. 
%The broken line designates the boundary of the adiabatic limit, defined when the maximum adiabatic parameter during a protocol obtains a value of $\mu=0.2$} 
%\end{figure}
\section{Eigenoperators existence proof}
\label{sec:proof of existence}
The form of Eq. \eqref{eq:factorized} seems to restrict the inertial theorem to specific Hamiltonian dynamics. This is not the case, as we will show in the following proof.
We claim that for any Hamiltonian $\hat{H}\b t$ with time-dependent analytical parameters one can find an orthonormal basis in Liouville space (a set of time-dependent orthonormal operators), such that, the generator of the dynamics in Liouville space can be decomposed as ${\cal{M}}\b t = {\cal P} {\cal{D}}\b t {\cal P}^{-1}$, where ${\cal P}$ is a constant unitary matrix and ${\cal D}\b t$ is a time-dependent diagonal matrix.

Proof: Let $\{\ket{\varphi}\}$ be an orthogonal basis of the system's $N$-dimensional Hilbert space. Then the set of matrices $\hat{X}_{nm} = \ket{\varphi_n}\bra{\varphi_m}$ form a basis for the associated $N^2$-dimensional Liouville space. The dynamics of the Liouville vector $\v{X}=\{\hat{X}_{1,1},\hat{X}_{2,1},...,\hat{X}_{N-1,N},\hat{X}_{N,N}\}$ is generated by the Heisenberg equation: $\f{d}{dt}\hat{X}_{j,k}^H\b t=i\sb{\hat{H}\b t,\hat{X}_{j,k}^H\b t}$. When the Hamiltonian is contained in the closed algebra, formed by $\{\hat{X}\}$, the dynamics of $\v X$ can be expressed as
\begin{equation}
    \f{d}{dt}\v{X}^H\b t = -i{\cal G}\b t \v{X}^H\b t~~.
    \label{eq:Xdynamics}
\end{equation}
Next, we define a new basis of Liouville space, given by $\v{F}^H\b t={\cal{V}}\b t\v{X}^H\b t$, where ${\cal V}$ is a time-dependent unitary transformation. Inserting the definition into Eq. \eqref{eq:Xdynamics} leads to
\begin{equation}
    \f{d}{dt} \v{F}^H\b t= -i{\cal{M}}\b t \v{F}^H\b t~~,
    \label{eq:proof1}
\end{equation}
where ${\cal{M}}\b t=-i{\cal{V}}\b{d{\cal{V}}^\dagger/dt}+i{\cal{V}}{\cal G}{\cal{V}}^\dagger$. The desired decomposition is now obtained by choosing a suitable basis transformation of the form ${\cal V}={\cal{P}}{{\cal{W}}}{\cal{O}}{\cal{P}}^{-1}$, where ${\cal P}$ is an arbitrary constant unitary matrix and ${{\cal W}}\b t$ is determined by the differential equation $d{\cal{W}}/dt =-i {\cal M}\b t{{\cal W}}$, with initial condition ${\cal W}\b 0=\cal{I}$, the identity in Liouville space. The matrix ${\cal O}$ is chosen to be a diagonal (in the $\{\hat{X}\}$ basis) time-dependent real matrix. Substituting ${\cal  V}$ into Eq. \eqref{eq:proof1} leads to the desired decomposition ${\cal M}\b t= {\cal P} {\cal{D}}\b t {\cal P}^{-1}$. 

To conclude, this result shows that for an analytic Hamiltonian, contained within the operator algebra, by choosing a suitable time-dependent operator basis, all the time-dependence of the generator can be absorbed into $\cal{D}$ and the time-dependent basis. Once such a basis is found, the inertial theorem can be used to generalize this specific solution to a broad family of inertial solutions.

%Here the $\lam_k$ and $\Omega_k$ correspond to parameters of ${\cal{D}}\b t$ in Sec. \ref{sec}
%The generator of Eq. \eqref{eq:proof1} should be  

\section{Combining inertial solutions and construction of inertial Hamiltonians}
\label{sec:construction_of_inertial_ham}
Constructing inertial solutions relies on the intuition related to the system's operator algebra. As a result, the solution may appear restrictive. This disadvantage can be overcome by combining inertial solutions to generate a new family of inertial protocols for a larger algebra. This  construction then generates a new inertial Hamiltonian. The method is based on the insight that deriving the Hamiltonian from a known propagator is relatively simple (in contrast, the reverse procedure, for a time-dependent Hamiltonian, is extremely involved).

We assume a set of known inertial solutions in Liouville space for the set of Hamiltonians $\{\hat{H}^{\b{i}}\b t\}$. The solutions are defined by the Liouville propagators $\{{\cal U}^{\b{i}}\b{t,0}\}$, these determine the dynamics of the operators basis set: $\v{v}^{H}\b{t}={\cal U}^{\b{i}}\b{t,0}\v{v}^{H}\b{0}$. Each Liouville propagator  has a matching propagator in the standard Hilbert space of wave-functions $\hat{U}^{\b{i}}\b{t,0}$. In turn, each propagator depends on rapid eigenvalues $\{\lambda^{\b{i}}\Omega^{{\b{i}}}\}$ and slowly varying inertial parameters $\{\chi^{\b{i}}\}$. By taking a product of inertial propagators we obtain a new inertial propagator:
\begin{equation}
    \hat{U}\b{t,0} = \hat{U}^{\b{1}}\b{t,0}\hat{U}^{\b{2}}\b{t,0}\cdots\hat{U}^{\b{n}}\b{t,0}~~,
    \label{eq:constructed_prop}
\end{equation}
which describes the dynamics accurately when all the inertial parameters change slowly.
The new inertial Hamiltonian is determined by the Shr\"odinger equation 
\begin{multline}
     \hat{H}\b t=i\pd{\hat{U}}{t}\hat{U}^{\dagger}=\hat{H}^{\b 1}+\hat{U}^{\b 1}\hat{H}^{\b 2}U^{\dagger\b 1}+\\
     \\\cdots+\hat{U}^{\b 1}\cdots \hat{U}^{\b{n-1}}\hat{H}^{\b n}\hat{U}^{\dagger\b{n-1}}\cdots \hat{U}^{\dagger\b 1} ~~,
    \label{eq:constructed_ham}
\end{multline}
where the time-dependence in the Hamiltonians and propagators is emitted to simplify the presentation.
The form of Eq. \eqref{eq:constructed_ham}, defines a broad family of inertial Hamiltonians, in each one the rapid degrees of freedom $\{\Omega\}$ can be set as arbitrary time-dependent continuous functions, while the inertial parameters satisfy inertial condition.

This procedure allows deriving inertial Hamiltonians for arbitrary dimensional systems. 
Moreover, since, Eq. \eqref{eq:constructed_prop} defines an inertial solution, the procedure can be repeated iteratively increasing the controlable degrees of freedom in the final Hamiltonian.

\subsection{Explicit demonstration}

To demonstrate the method we combine two inertial solutions of the two-level system, Sec. \ref{sec:TLS1}, and derive a three-level inertial Hamiltonian. The first step requires mapping the Liouville propagator to the associated Hilbert space propagator. The transformation is performed by utilizing the vec-ing procedure, which maps the Hilbert space dynamics to Liouville space \cite{fano1957description,scopa2019exact}. The density operator maps to a vector in Liouville space $\v{r}$  by ordering the columns of $\hat{\rho}$, i.e., for an $N$ by $N$ density operator the $\b{i,j}$ matrix entry of $\hat{\rho}$ maps to the $\b{j-1}N+i$ entry of $\v{r}$. Operation of operators from both left and right on $\hat{\rho}$, as $\hat{A}\hat{\rho}\hat{B}$, maps to the operation $\b{\hat{B}^T\otimes\hat{A}}\v r $  in Liouville space. Therefore, the time evolution $\hat{U}\b{t,0}\hat{\rho}\b 0\hat{U}^\dagger\b{t,0}$ leads to the Liouville propagator ${\cal{U}}\b{t,0}=\hat{U}^{*}\b{t,0}\otimes\hat{U}\b{t,0}$, where $^*$ designates the complex conjugation operation. 

In the eigenstates representation $\{\ket{\phi_k\b t}\}$ the wave-function propagator can be expressed as
\begin{equation}
    \hat{U}\b{t,0}=\sum_k^N e^{i\alpha_k\b t}\ket{\phi_k\b t}\bra{\phi_k\b 0}~~,
    \label{eq:U_hilbert}
\end{equation} 
where $\{\alpha_k\}$ are the associated phases.
Transforming the dynamics according to the vec-ing procedure in the eigenbasis representation of $\hat{U}\b{t,0}$ leads to ${\cal U}\b{t,0}=\text{diag}\b{\sum_{k,l}^N e^{i\b{\alpha_k\b t-\alpha_l\b t}}}$, while $\v r$ corresponds to the Liouville state vector in the time-dependent operator basis $\{\ket{\phi_k\b t}\bra{\phi_l\b t}\}$. The entries of $\v r \b t$  are given by the matrix element of $\hat{\rho}\b t$ in the $\{\ket{\phi_k\b t}\}$ basis. 

In Liouville space the propagator is diagonal in the eigenoperator basis, which implies that the set of eigenoperators correspond to the set $\{\ket{\phi_k\b t}\bra{\phi_l\b t}\}$. This identification allows deriving the eigenstates of $\hat{U}\b{t,0}$ and the matching phases $\{\alpha_k\b t\}$, which determine the propagator, Eq. \eqref{eq:U_hilbert}.

We demonstrate the reversed mapping by deriving the Hilbert space propagator for the inertial solution of the two-level system, Sec. \ref{sec:TLS1}. By following a similar methodology any Liouville propagator can be mapped to the corresponding Hilbert space propagator. 
First, the eigenoperators of the propagator $\{\hat{F}_1,\hat{F}_2,\hat{F}_3\}$ are obtained diagonalizing the generator $\bar{{\cal{B}}}$, Eq. \eqref{TLS propagator}. Formally, the eigenoperators are the operators associated with the rows of the diagonalizing matrix ${\cal{P}}^{-1}$, satisfying ${\cal{P}}\bar{{\cal{B}}}{\cal{P}}^{-1}=\text{diag}\b{0,\bar{\kappa},-\bar{\kappa}}$. We next normalize these Liouville vectors to obtain an orthonormal set of  corresponding operators $\{\hat{G}_1,\hat{G}_2,\hat{G}_3\}$. These are characterized by simple dynamics (Heisenberg picture): $\hat{G}_1^{H}\b t=\hat{G}_1\b 0$, $\hat{G}_2^{H}\b t=e^{-i \int_0^t\bar \kappa\bar\Omega dt'}\hat{G}_2\b 0$ and $\hat{G}_3^{H}\b t=\b{\hat{G}_2^{H}\b{t}}^\dagger$. The operators  $\hat{G}_2\b t\hat{G}_3\b t$ and $\hat{G}_3\b t\hat{G}_2\b t$ are projection operators with common instantaneous eigenstates (in Hilbert space)  $\{\ket{\phi_k\b t}\}$ (the eigenstates of $\hat{U}\b{t,0}$) with eigenvalues $\{1,0\}$. Finally, the mapping to the Hilbert space propagator is achieved by calculating the expectation values of $\{\hat{G}_i\b t\}$, expressing the operators in terms of the  eigenstates  at time $t$ and the initial state  $\hat{\rho}\b 0$ in terms of $\{\ket{\phi_k\b 0}\}$. The expectation value of $\hat{G}_2\b t$ then becomes (see Appendix \ref{apsec:generalization} for further details) % \ref{apsec:generalization}
\begin{equation}
    \bra{\phi_1\b 0}\hat{U}^{\dagger}\b{t,0}\ket{\phi_1\b t}\bra{\phi_2\b t}\hat{U}\b{t,0}\ket{\phi_2\b 0}=e^{-i\Lambda\b t}~~,
    \label{eq:exp_G2}
\end{equation}
where $\Lambda\b t=\int_0^t\bar \kappa\bar\Omega dt'+\bar{\Delta}/2$, with $\bar{\Delta}=\lambda\b{t}-\lambda\b{0}$ and $\lambda\b{t}=\f{i\eps\bar \kappa+\bar \mu\omega}{\sqrt{\eps^{2}\bar \kappa^{2}+\bar \mu^{2}\omega^{2}}}$. Similarly, the expectation value of $\hat{G}_3\b t$ leads to the conjugate equation of Eq. \eqref{eq:exp_G2}. These identities imply the explicit form of the Hilbert space propagator: 
\begin{equation}
    \hat{U}\b{t,0}=e^{i\Lambda\b t}\ket{\phi_{1}\b t}\bra{\phi_{1}\b 0}+e^{-i\Lambda\b t}\ket{\phi_{2}\b t}\bra{\phi_{2}\b 0}~~.
    \label{eq:propagator_hilb}
\end{equation}

For different inertial solutions, such as the harmonic oscillator or three-level system Secs. \ref{subsec: paremetric HO} and \ref{subsec:3 level}, the Hilbert space propagator is obtained by following a similar procedure. First the eigenoperators are divided to pairs, each constitute adjoint pairs. Taking a product of the two operators (two options exist $\hat{G}_i\hat{G}_j$ and $\hat{G}_j\hat{G}_i$) gives a projection operator of an eigenstate $\ket{\phi_k\b t}$. The eigenstates can then be obtained via diagonalization. The two corresponding eigenstates lead to an analogous equation as Eq. \eqref{eq:exp_G2}, which determines the propagator phases $\alpha_{k}\b t$ and subsequently, from Eq. \eqref{eq:U_hilbert}, the Hilbert space propagator.

We have now reached the point where we can combine two TLS propagators $\hat{U}^{\b{1}}\b{t,0}$ and $\hat{U}^{\b{2}}\b{t,0}$, Eq. \eqref{eq:propagator_hilb}, to obtain an inertial Hamiltonian for a three-level system $\hat{U}\b{t,0}=\hat{U}^{\b{1}}\b{t,0}\hat{U}^{\b{2}}\b{t,0}$. We consider two propagators that couple two energy levels with a single common energy state, which is taken to be the second energy state. Each propagator is generated by an Hamiltonian in the form of Eq. \eqref{eq:TLS hamil}, with distinct frequencies $\eps^{\b{i}}\b{t}$, $\omega^{\b{i}}\b{t}$ and inertial parameters $\bar \mu^{\b{i}}\b{t}$, with $i=1,2$. Substituting the associated propagators, Eq. \eqref{eq:propagator_hilb}, into Eq. \eqref{eq:constructed_ham} leads to a new inertial Hamiltonian
\begin{multline}
    \hat{H}\b t = \hat{H}^{\b{1}}\b{t}+\hat{H}^{\b{2}}\b{t}+\hat{H}_{12}\b{t}~~,
    \label{eq:constructed_ham_example}
\end{multline}
where the correction term is
\begin{equation}
 \hat{H}_{12}\b t = \sb{\hat{U}^{\b{1}}\b{t,0},\hat{H}^{\b{2}}\b{t}}\hat{U}^{\b{1}\dagger}\b{t,0}   
    \label{eq:H_12}
\end{equation}
and $\sb{\hat{U}^{\b{1}},\hat{H}^{\b{2}}}$ is given in Appendix \ref{apsec:generalization}. %\ref{apsec:generalization} 

The norm of the correction Hamiltonian $\hat{H}_{12}$ depends on the control parameters 
$\hat{U}^{\b{1}}$ and $\hat{U}^{\b 2}$. We compared the norm of the correction to the norms of  $\hat{H}^{\b 1}$ and $\hat{H}^{\b 2}$ and find that for typical parameter values the norm is at 
least an order of magnitude smaller ($|\mu|\sim 1$ and $\Omega^{\b{i}}\sim 10$) . This suggests a perturbative approach with respect to this correction.

\section{Extending the inertial theorem to open-system dynamics}
\label{sec:open system}
The inertial solution describes the free propagation of isolated systems. In reality no system is truly isolated, as a consequence, the environment modifies  the system dynamics. By combining the inertial theorem and the Non-Adiabatic Master Equation (NAME)  \cite{dann2018time}
 a reduced description of the system dynamics can be obtained, where the influence of the bath is treated implicitly.

The crucial step of the derivation includes solving the free propagation, which in turn is used to obtain the system-bath interaction Hamiltonian $\hat{H}_I$ in the interaction representation. Applying the inertial theorem to expand $\hat{H}_I$ in terms of the eigenoperators $\hat{F}_k$, Eq. \eqref{eq:inetrial state} and following the derivation presented in \cite{dann2018time} leads to a NAME incorporating the effect of a slowly accelerated drive, $d\v \chi/dt=d \mu/dt>0$ (see Appendix \ref{ap:open system}).%\ref{ap:open system}

The Master equation in the interaction representation is given
\begin{multline}
\f{d}{dt} \tilde{\rho}_S \b t= -i\sb{\tilde{H}_{LS}\b t,\tilde{\rho}_S\b t}\\+ \sum_{j} \gamma_k\b{ \alpha_k\b t} \b{\hat{F}_k \tilde{\rho}_S\b t \hat{F}_k^{\dagger} -\f{1}{2}\{\hat{F}_k^{\dagger} \hat{F}_k \tilde{\rho}_S\b t\}}~~.
\label{non_adiabatic_master_eq}
\end{multline}
Here, $\tilde{\rho}_S\b t$ is the system's density operator in the interaction representation relative to the free evolution, and  $\hat{F}_{j} \equiv \hat{F}_{j} \b 0$.  The term $\tilde{H}_{LS}\b t$ is the time-dependent Lamb-type shift Hamiltonian.  This Master equation, Eq. \eqref{non_adiabatic_master_eq}, is an explicit time-dependent version of the Markovian Gorini-Kossakowski-Lindblad-Sudarshan (GKLS) Master equation  \cite{lindblad1976generators,gorini1976completely}.

Within the derivation of Eq. \eqref{non_adiabatic_master_eq}, the inertial theorem eigenoperators, $\hat{F}_k$, Eq. \eqref{eq:inetrial state}, are identified as the jump operators of the Master equation. These determine the instantaneous attractor of the dynamical map and the decay rates \cite{dann2018time}.
The decay rates $\gamma\b{\alpha_k}$ are related to the Fourier transform of the bath correlation functions with effective frequencies $\alpha_k\b t$. These effective frequencies are the derivative of the accumulated phases, associated with the eigenvalues of $\hat F_k$.
In Appendix \ref{ap:open system}, the construction of Eq. \eqref{non_adiabatic_master_eq} is demonstrated for a driven system weakly coupled to a bath.  %\ref{ap:open system}

The framework of the inertial NAME, Eq. \eqref{non_adiabatic_master_eq}, has been employed in an open system control problem: accelerating thermalization \cite{dann2019shortcut,dann2020fast}. In addition, the theory allowed constructing a fully quantum Carnot analog engine \cite{dann2020quantum,dann2020quantum2}, highlighting quantum signatures in quantum thermodynamics \cite{uzdin2015equivalence,miller2019work}.

\section{Geometric phase in Liouville space}
\label{sec:geometric phase}
%\tb{gaps: something on the dimension of the berry phase maybe, and an example of a 2D problem, and a proof that for a non-closed circuit the berry phase is of second order. }
In 1984, Berry showed that a system transported adiabatically, by varying parameters of the Hamiltonian, around a circuit, acquires an additional geometric phase \cite{berry1984quantal}.
Following a similar proof, we show that the operator $\hat{F}_k \b{\theta}$ attains a new geometric phase, $\phi_k$, when the parameters $\{ \chi \}$ are transformed slowly in circuit $C$ in parameter space, Cf. Appendix \ref{geometric proof}. %\ref{geometric proof}
The geometric phase has the form
\begin{equation}
\phi_{k}\b{C}=-\text{Im} \sb{\iint_{C}d\v \chi\cdot {\cal{V}}_k \b{\v{\chi}}
}~~,
\label{eq:Gamma}
\end{equation}
where 
\begin{equation}
{\cal{V}}_k \b{\v{\chi}} = \sum_{n\neq k}\f{\b{\v G_{k},\v{\nabla}_{\v{\chi}}{\cal M}\v F_{n}}\times\b{\v G_{n},\v{\nabla}_{\v{\chi}}{\cal M}\v F_{k}}}{\b{\Omega_n\lam_{n}-\Omega_k\lam_{k}}^{2}}~~.
\end{equation}

The geometric phase in Liouville space, Eq. \eqref{eq:Gamma}, has a different physical significance compared to the Berry phase. The Berry's phase is an accumulated phase of the wave-function, and therefore, is non-vanishing only for a closed circuits including a degeneracy of eigen-energies. As a property of the wave-function, it can only be witnessed by interference.

In contrast, the geometric phase in Liouville space influences the physical observables directly. Such observables are determined by a linear combination of the eigen-operators, (associated with the vector $\v v^H$, Eq. \eqref{eq:inetrial state}) and the initial system state. Moreover, unlike the Berry phase, the eigenvectors $\v{F}_k$ are uniquely defined by $\v\chi$. As a result, $\phi_k$ is non-vanishing for open circuits in the parameter space $\{\chi\}$. For a closed circuit, $\phi_k$ vanishes when ${\cal{V}}_k \b{\v{\chi}}$ is analytic within the area encompassed by the circuit. When the circuit surrounds a singularity, which may occur in the case of degeneracies $\Omega_n\lam_{n}=\Omega_k\lam_{k}$, the geometric phase is non-vanishing and will directly affect the physical state. 

\subsection{Geometric phase examples in Liouville space}

As a first demonstration we consider a two-level-system in a time-dependent magnetic field. This system is represented by the Hamiltonian
\begin{equation}
    \hat{H}\b t= \bar{\Omega}\b{B_x\b t \hat{S}_x +B_y\b t \hat{S}_y +B_z\b t \hat{S}_z }~~,
    \label{eq:geom Ham}
\end{equation}
where $B_i\b t=f\b t b_i\b t $ are the components of the magnetic field $\v B\b t$ and  $f\b t=|\v B|$ is a time-dependent function. In terms of the spin operator basis $\{\hat{S}_x,\hat{S}_y,\hat{S}_z\}$, the dynamics in Liouville space are generated by
\begin{equation}
    \f d{dt}\v S^H\b t=-i{\cal H}\b t\v S^H\b t~~,\\
\end{equation}
with ${\cal{H}}=i\bar{\Omega}\{{0,-B_z,B_y};{B_z,0,-B_x};{-B_y,B_x,0}\}$.
The required decomposition of the dynamical equation in Liouville space, Eq. \eqref{eq:factorized}, is obtained for slowly varying $b_i\b t$. For such Hamiltonian, the inertial limit coincides with the adiabatic limit under the scaling $t\ra\int_0^t\,dt'f\b{t'}$. Nevertheless, this example is instructive, as it demonstrates the properties of the geometric phase in Liouville space, and highlights the distinctions relative to the Berry phase in the wave-function Hilbert space. 

\begin{figure}[htb!]
\centering
\includegraphics[width=7cm]{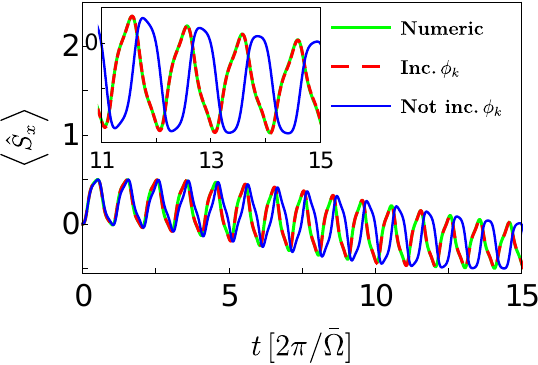}%Sx_geom_TLS.pdf
\caption{ Expectation value of $\hat{S}_x$ as a function of time, for an converged numerical solution (green), complete inertial solution (red dashed) and an inertial solution not including the geometric phase (blue). The inset emphasises the difference in phase at long times. The error in phase accumulates with time, while the complete inertial solution follows the converged solution faithfully. Model parameters are: $f\b t = 2+\sin\b{4 t}$, $\bar{\Omega}=2$, $\theta=3\pi/4$ and the system is initialized in the ground state $\v v \b 0=\{ 0,0,-\hbar/2\}$ (in the spin operator basis). The total accumulated geometric phase is $\cos\b{\theta}\Delta\phi =\sqrt{2}\pi$. } %$\phi_{tot}=\cos\b{\theta}\Delta\varphi=\sqrt\b 2 \pi$. }
\label{fig:Sx_geom_TLS}
\end{figure}

\begin{figure}[htb!]
\centering
\includegraphics[width=7cm]{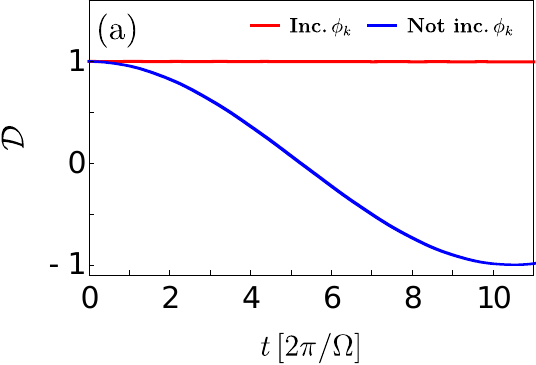}\\%geom_dist.pdf
\includegraphics[width=7cm]{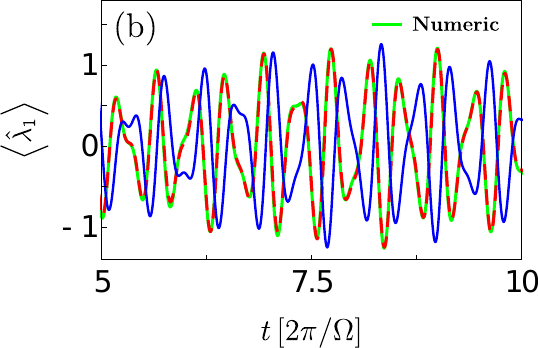}%geom_lam_1.pdf\\
\caption{\label{fig:geometric_SU3} Geometric phase for a three-level-system, Eq. \eqref{eq:3level_geom}. Panel (a): The distance ${\cal{D}}\b{\v u,\v v}\equiv\b{\v u , \v v}$ between the inertial solution relative to a converged numerical solution as a function of time. The red line includes the geometric phase while the blue thin line does not (both solutions include the dynamical phase).  Panel (b): The expectation value of the first Gell-Mann operator $\hat{\lam}_1$ as a function of time.
As expected, the distance between the converged solution and the solution not including the geometric phase increases. In contrast the complete inertial solution remains accurate, achieving distances close to unity. All the system observables show a typical behaviour as presented in Panel (b). The observable of the complete inertial solution remains close to the numerical solution, while the solution lacking the geometric phase deviates from the expected value as the phase increases. Gaps between the inertial and numerical solution arise from the small deviation from the inertial limit. The model parameters are: $\Omega =1/\sqrt{2}$, $\Delta=-\sqrt{2}$, $\chi=0.75$, $\xi=0.5$, $\alpha\b t=-\Omega\sin\b{\Omega \chi t}$, $\beta\b t=\Omega\cos \b{\Omega \chi t}$, $\eta_1 =\cos\b{g\b t}$ and $\eta_2 = \sin\b{g\b t}$  with $g\b t = \b{7\pi/4}\b{t/t_f}+ \pi/4$ for a final time $t_f \text{a.u.}$. For a complete circle in parameter space, the total accumulated phase is $\phi_{tot} \approx 1.04 \pi$. } 
\end{figure}

For simplicity, the magnetic field is varied while keeping the magnitude of $\v b\b t=\{b_x\b t,b_y\b t,b_z\b t\}^{T}$ constant and equal to unity. For such a protocol, it is natural to express the dynamics in terms of the spherical angles ($\v{b}\b{t}=\v{b}\b{\theta\b t,\varphi\b t}$), leading to the eigenoperators $\hat{F}_k\b{\theta,\varphi}$ and associated geometric phases in Liouville space $\phi_k$, see Appendix \ref{ap:geo example}. A rotation of  $\v b \b t$ towards direction $\hat{n}\b{\theta_f,\varphi_f}$ leads to accumulated geometric phases: $\phi_{\pm}=\pm \cos\b{\theta_f}\Delta\phi$, where $\Delta \varphi=\varphi_f-\varphi_i$ is the change in the azimuthal angle, associated with eigenoperators $\hat{F}_\pm$ correspondingly. These induce a direct effect on the system state. %\ref{ap:geo example}

%\begin{figure}[htb!]
%\centering
%\includegraphics[width=6.5cm]{3D-1.pdf}
%\caption{Landscape of eigenvalues $\{\lam\}$ of $\cal M$ as a function of $\eta_1$ and $\eta_2$, Eq. \eqref{eq:3level_geom}. %The green and yellow rings present the studied protocol, initializing in the state $\hat{\rho}\b %0=\f{1}{3}\hat{I}+\hat{v}_1\b 0+\hat{v}_8\b 0$, where $\hat{v}_{1/8}$ are the eigenoperator corresponding to $\lam_{1/8}$. %The protocol parameters are stated in Fig. \ref{fig:geometric_SU3}.  }
%\label{fig:geom_surface}
%\end{figure}

In Fig. \ref{fig:Sx_geom_TLS} we compare the inertial solution, including the geometric phase, to a solution where the phase is omitted, and an converged numerical solution. The result demonstrates the importance of the geometric phase for an accurate dynamical description. Moreover, unlike the Berry phases in Hilbert space, $\phi_k$ influence the dynamics for closed as well as open circuits in the $\b{B_x,B_y,B_z}$ space.

An addition example demonstrating the geometric phase employs the Hamiltonian of a fully driven three-level system:
 \begin{equation}
     \hat{H}\b t=\sb{\begin{array}{ccc}
\frac{\Delta\alpha^{2}+2\beta^{2}\delta}{2\Omega^{2}} & \frac{\b{\eta_{1}-i\eta_{2}}\alpha}{\Omega} & \frac{\alpha\beta\b{\Delta-2\delta}}{2\Omega^{2}}\\
\frac{\b{\eta_{1}+i\eta_{2}}\alpha}{\Omega} & -\frac{\Delta}{2} & \frac{\b{\eta_{1}+i\eta_{2}}\beta}{\Omega}\\
\frac{\alpha\beta\b{\Delta-2\delta}}{2\Omega^{2}} & \frac{\b{\eta_{1}-i\eta_{2}}\beta}{\Omega} & \frac{2\delta\alpha^{2}+\beta^{2}\Delta}{2\Omega^{2}}
\end{array}}~~,
\label{eq:3level_geom}
 \end{equation}
 where all the Hamiltonian parameters can be time-dependent and $\Omega=\sqrt{\alpha^2+\beta^2}$.
 The inertial solution for Hamiltonian Eq. \eqref{eq:3level_geom} is derived in a similar manner as in Sec. \ref{subsec:3 level}.  The time-dependent transformation $\hat{S}$, Eq. \eqref{eq:strans}, leads to the dynamical operator basis in Liouville space and the $\cal M$ matrix. The associated inertial variables are: $\chi=\b{\alpha\dot{\beta}-\dot{\alpha}\beta}/\Omega^3$, $\gamma=\Delta/\Omega$, $\eta_1$, $\eta_2$ and $\xi=\delta/\Omega$.
% Figure \ref{fig:geom_surface} presents the manifolds of the eigenvalues of $\cal M$ (of the Hamiltonian of Eq. \eqref{eq:3level_geom}). The two rings present an inertial protocol in parameter space, employed to demonstrate the geometric phase.  

An excursion on the eigenvalue manifold is generated by varying the parameters $\eta_1$ and $\eta_2$.
%, which is shown in Fig. \ref{fig:geom_surface}.
Figure 
\ref{fig:geometric_SU3} compares a numerical solution of the dynamics with the inertial solution, with and without the geometric phase. We take the initial state as $\hat{\rho}\b 0=\f{1}{3}\hat{I}+\hat{v}_1\b 0+\hat{v}_8\b 0$, and vary $\eta_1$ and $\eta_2$ cyclically. The errors of the complete inertial solution arise from deviations from the inertial limit.

The overall phase of the inertial solution is of importance in the derivation of open system dynamics Cf. \ref{sec:open system}. The phase enters into the detailed balance expression, and thus, is influenced by the geometric phase. As a result, the kinetic rates of the Master equation are dependent on the geometric phase.

%We preform a rotation of $\b{b}\b t$ while varying $f\b t$ rapidly according to a linear protocol and compare the results with an converged numerical solution. The results are presented in 

%Here, $\hat{H}^i \b t= \omega_i \b t \hat{S}_z^i+\eps_i \b t \hat{S}_x^i$ with $i=1,2$. The driving is of the form, $\omega_i \b t = \bar{\Omega} \b t \text{cos}\b{ \alpha_i\b t} $  and $\eps_i\b t= \bar{\Omega} \b t \text{sin}\b{ \alpha_i\b t }$, where both spins have identical Rabi frequencies, $\bar \Omega_1=\bar \Omega_2=\bar{\Omega} \b t$, and $\alpha_i \b t$ is determined by the parameter $\chi_i$, Cf. SI Appendix \ref{ap:geo example}.   

%The dynamics of such a system can be cast in a factorized form, Eq. \eqref{theta_motion}, which includes two independent constant parameters $\chi_1$ and $\chi_2$ (see SI Appendix \ref{ap:geo example} Eq., Eq. \eqref{r_nl eq}). The analysis of the dynamics leads to the conclusion that  non local operators, such as $\hat{L}_1\otimes\hat{L}_2$,  are affected by the geometric phases that originate from a trajectory in the parameter space of $\chi_1$ and $\chi_2$.

\section{Discussion}
\label{sec:discussion}
To summarize, the inertial solutions are constructed in Liouville space, employing a time-dependent operator basis. The transformation to this basis incorporates a part of the time-dependence, while, an additional fast timescale is absorbed into the eigenvalues of the propagator $\{\Omega\b t \lam\b{\v \chi}\}$. This leads to the required decomposition, allowing to bypass the time-ordering obstacle.
The  transformation is the key towards obtaining a suitable decomposition, leading to the inertial solution.
An appropriate time-dependent operators basis always exists (Cf. Sec. \ref{sec:proof of existence}) but its construction requires ingenuity.
The symmetries of Lie algebras can serve as a guideline to guess an appropriate transformation to the time-dependent basis.

%A time-dependent basis which allows
%the construction (3) can always be found for the trivial case of M(t)
%being the identity operator. Ingenuity is required when an M(t) (or
%D(t)) is sought which has a sufficiently large gap, allowing the
%stability advantages of the method to be realized.

The solution is characterised by a set of slowly varying inertial variables $\{\chi \}$, and expressed in terms of the eigenoperators $\{\hat{F}\}$, along with dynamical and geometric phases $\{\phi\}$. 
%These eigenoperators, up to a global scaling , represent time-dependent invariants.
The eigenoperators with vanishing eigenvalues correspond to time-dependent constants of motion \cite{wulfan1981adiabatic,pfeifer1983stationary}. Meaning that the expectation values of these operators are invariant under the driven dynamics.  In contrast, the eigenoperators with non-vanishing eigenvalues carry a time-dependent phase.

The inertial solutions share common features with the quantum adiabatic solutions. These are a consequence of the quantum adiabatic theorem, which was derived ninety years ago by Born and Fock \cite{born1928beweis}.
It states that for a slowly varying Hamiltonian $\hat H \b t$, an eigenstate $\ket{n\b 0}$ of the initial Hamiltonian $\hat{H}\b 0$, remains an eigenstate $\ket{n \b t}$ of the instantaneous Hamiltonian $\hat{H}\b t$ throughout the process. 
By inserting the instantaneous eigenstate solution into the time-dependent Schr\"odinger equation, the validity of the adiabatic approximation is be determined.

Comparing the structures of the two solutions, the eigenstates of the instantaneous Hamiltonian in the adiabatic solution are replaced by the eigenoperators of the propagator in the inertial solution. The dynamics in both cases includes a geometric and dynamical phases, which differ by their physical meaning. Moreover, the validity of the inertial solution is verified in a similar fashion, by substituting the solution into the dynamical equation in Liouville space. In the special case, where a single transformation generates the time-dependent dynamical operator basis from a time-independent basis, the inertial solution can be formulated as an adiabatic solution in the interaction representation in Hilbert space.

The differences between the adiabatic and inertial solutions are highlighted by applying both methods to the same driven system. This is quantified by the inertial and adiabatic parameters, $\Upsilon$ and $\mu$. The adiabatic parameter restricts the range of validity of the adiabatic approximation, while the inertial parameter limits the inertial solutions. 
Typically, the inertial parameter is associated with the change in the adiabatic parameter, and therefore,  the range of valid protocols is significantly enhanced to include rapid changes associated with a large $\mu$. 

% These phases influence the system observables directly, and can be significant under closed as well as non-closed circuits in the inertial variables space.
%In all cases tested, the inertial solution shows high accuracy with respect to the converged numerical solution.

%\tg{Another important requirement in quantum control is the stability of the solution, under variations in the driving protocol and external noise. }

Corrections to the adiabatic approximations have been developed.
There is some confusion in their terminology. One approach  is based on adding a counter-diabatic control to the Hamiltonian, which eliminates transition terms between adiabatic states, thus, maintaining adiabatic evolution \cite{demirplak2003adiabatic,lim1991superadiabatic,sels2017minimizing}.  
A different approach  termed super-adiabatic is based on a perturbative treatment in orders of the adiabatic parameter \cite{lim1991superadiabatic,berry1990histories}. 
An alternative is to employ the adiabatic transformation iteratively.   
For the $SU(2)$ case  this can be carried out to infinite order \cite{demirplak2008consistency,ibanez2012multiple}. This approach could be hard to generalize for a larger algebra with more than one adiabatic parameter and is closest in spirit to the inertial procedure.

%\tg{The inertial theorem generates a new family of solutions, which are not corrections to the adiabatic solution, and do not require additional control operators. These solutions enable rapid control tasks. Additional improvements to the inertial solutions can be applied, in a similar fashion to the adiabatic corrections: counter-inertial control or super-inertial basis.}
%\tg{In addition, a numerical approach to obtain solutions based on the adiabatic limit has been developed \cite{jahnke2003numerical}. This method incorporates the idea of timescale separation. A similar approach could be employed based on the inertial solutions.}

Generally, similar closed-form solutions serve as a platform to construct control protocols. For example, the adiabatic STIRAP procedure, which has become extremely popular in contemporary physics \cite{vitanov2017stimulated}. As closed-form solutions, the inertial solutions also generate a constructive family of control strategies. As  an example we introduced the inertial STIRAP, which can be incorporated in similar control tasks as the adiabatic STIRAP, but allow for faster control. 

An additional possible control strategy concerns the geometric and dynamical phases of the inertial solution. These  serve as a new template for interference. Overall, since the inertial solution is stable \cite{hu2019experimental}, the control protocols, based on the solution, are expected to be robust in the presence of noise \cite{childs2001robustness}. An analysis of the noise has been performed in  Eq. \eqref{noise E of M} for the three-level system. The generalization to other algebras is straight forward. Similar treatment concerning the robustness were carried out for the Landau-Zener scenario  \cite{kayanuma1984nonadiabatic,pokrovsky2003fast}.

The scope of inertial solutions can be extended by combining individual inertial solutions, Sec. \ref{sec:construction_of_inertial_ham}. For example, in a multilevel Hilbert space inertial solutions with a common levels can be combined, generating a new time-dependent control Hamiltonian.

The inertial solution also has a direct application for open-system control. Since any quantum system interacts with the environment to some extent, complete control includes taking into account the affect of the environment. In contrast to the typical analysis, the driving has a direct influence on the system-bath interactions \cite{dann2018time}. In the weak system-bath coupling limit, the influence of the free dynamics can be incorporated by the Non-adiabatic Master equation, Sec. \ref{sec:open system}. In such open system control scenarios, the controller affects the system directly through the driving and indirectly through the system-bath coupling. This property enables extending the common unitary quantum control to transformations involving changes in entropy \cite{dann2019shortcut}.

\section{Acknowledgement}
We thank KITP for their hospitality and support. This research was supported by the Adams Fellowship  Program of the Israel Academy of Sciences and Humanities, the National Science Foundation under Grant No. NSF PHY-1748958 and The Israel Science Foundation Grant No.  2244/14. We thank David Tannor, Raam Uzdin, Amikam Levy, Erik Torrontegui, Moshe Armon and Marcel Fabian for fruitful discussions.
\appendix
\section{Geometric phase}
\label{geometric proof}
We derive the geometric phase in Liouville space, assuming a general non-hermitian generator $\cal{M}$, (Eq. (3) in the main text). The derivation follows the original derivation of Berry \cite{berry1984quantal}, extending the solution to a non-hermitian generator.
When $\v v^H\b{\chi\b t}$ completes a contour $C$ in the parametric  of $\{ \chi \}$ (not necessarily a closed), the inertial solution acquires a geometric phase of the form
\begin{equation}
\phi_{k}\b{\v \chi\b t}=i\ointop_C d\v{\chi}\cdot\b{\v{G}_{k},\nabla_{\v{\chi}}\v F_{k}}~~.
\end{equation}

%\trr{discuss the dimensions of the problem and the suitable way to solve for them.}
  
When the matrix $\cal{M}$ includes three inertial variables $\v{\chi}=\{\chi_1,\chi_2,\chi_3\}^T$, the calculation of the geometric phase is simplified by utilizing common vector calculus identities and Stoke's theorem. Following Berry's derivation \cite{berry1984quantal}, and identities Eq. (11) and Eq. (12) in the main text lead to the final result
\begin{multline}
\phi_{k}\b{\theta}=\\
\text{Im}\sb{\iint_{C}d\v s\cdot\sum_{n\neq k}\f{\b{\v{G}_{k},\nabla_{\v{\chi}}{\cal M}\v F_{n}}\times\b{\v{G}_{n},\nabla_{\v{\chi}}{\cal M}\v F_{k}}}{\b{\Omega_n\lam_{n}-\Omega_k\lam_{k}}^{2}}}~~.
\end{multline}

\section{Comparison of the inertial and adiabatic solutions for slow driving}
We compare the inertial and adiabatic solutions to  converged numerical results for slow driving. A linear ramp protocol is considered, for the harmonic oscillator model the oscillator frequency increases linearly in time 
\begin{equation}
\omega \b t = \omega \b 0 +\delta t~~,
\end{equation}
with $\delta = \b{\omega\b{t_f}-\omega\b 0}/t_f$.
The two-level-system is modified by a similar protocol, varying the Rabi frequency linearly, $\bar{\Omega} \b t =\bar{\Omega} \b 0 +\bar{\delta} \cdot t$.

\begin{figure}[htb!]
\centering
\includegraphics[width=8cm]{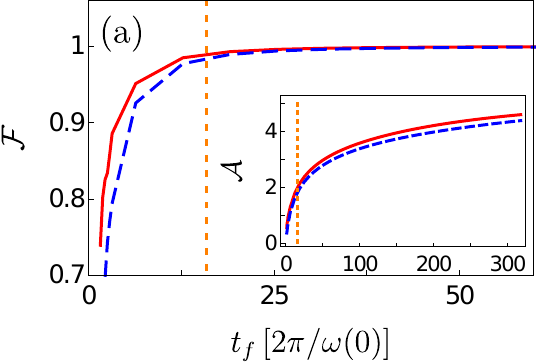}\\%fidelity_lin_HO.eps
\vspace{1cm}
\includegraphics[width=8cm]{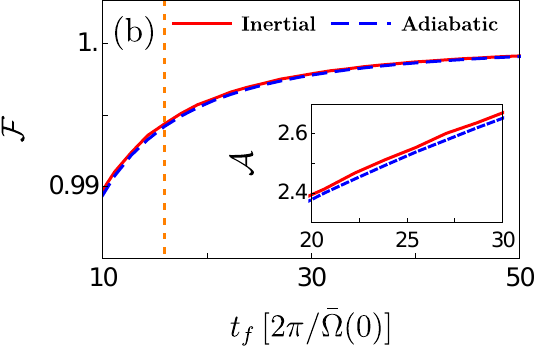}\\%fidelity_lin_TLS2.eps
\caption{\label{fig:slow driving} The fidelity of the final state as function of the protocol time $t_f$ for the (a) harmonic oscillator and (b) two-level-system. (Inset) Results shown at a higher resolution, in the adiabatic regime (${\cal{A}}=-\log\b{1-{\cal{F}}}$). The dashed vertical line designates the boundary of the adiabatic limit, defined when the maximum adiabatic parameter during a protocol obtains a value of $\mu=\bar{\mu}=0.2$. As the accuracy improves the fidelity converges to unity and $\cal{A}$ increases. The inertial solution has a superior accuracy in both the adiabatic (right of the vertical dashed line) and non-adiabatic (left of the vertical dashed line) regimes.
Calculation parameters for the HO are:
$\omega\b 0=10$, $\omega\b{t_f}=100$.
The parameters for the TLS are:  $\bar{\Omega} \b 0 = 10$, $\bar{\Omega} \b{t_f}= 100$ for $\eps=8$ with initial values $\mean{\tilde{H}\b 0}=4$ and $\mean{\tilde{L}\b 0}=\mean{\tilde{C}\b 0}=1$.} 
\end{figure}

The comparison between the two solutions is presented in Fig. \ref{fig:slow driving}, demonstrating the superiority of the inertial approximation over the adiabatic approximation, in both the adiabatic and non-adiabatic regimes.

\section{Non-Adiabatic Master Equation (NAME)}
\label{ap:open system}
We present a derivation of a Master equation for a driven quantum system interacting with a thermal electromagnetic field with temperature $T$, see Ref. \cite{dann2018time} for further analysis. \\
The composite system is represented by the Hamiltonian
\begin{equation}
\hat H_{tot}\b t=\hat H\b t+\hat H_B+\hat H_I~~,
\end{equation}
where $\hat H\b t$ is the driven system Hamiltonian, the bath Hamiltonian is composed of all the bath modes of the form $\hat H_B=\sum_{\lam=1,2}\sum_{\v{k}}\omega_k \hat b_{\lam}^{\dagger}\brac{\v{k}}\hat b_{\lam}\brac{\v{k}}$, and $\hat H_I$ is the system bath interaction term.
The interaction term under the dipole approximation can be written as $\hat H_I=\v{E}\cdot \v{D}$, where
$\v{D}$ is the system dipole operator and $\v{E}$ is the electromagnetic field operator. Such a field operator obtains the form $\v{E}=i\sum_{\v{k}}\sum_{\lambda=1,2}\sqrt{\f{ \hbar \omega_k}{2 \eps_0 V}}\v{e}_k \brac{\v{k}}\b{\hat b_k\brac{\v{k}}+\hat b_k^{\dagger}\brac{\v{k}}}$, where $V$ is the volume of the field, $\eps_0$ is the electric constant, $\hat b_\lam\brac{\v{k}}$ and $\hat b_{\lam}^{\dagger}\brac{\v{k}}$ are the annihilation and creation operators of a bath mode in the $\hat{k}$'th direction with a frequency $\omega_k$, ($k\equiv|\v k|$), and polarization $\lambda$.

Following the microscopic derivation \cite{breuer2002theory,alicki2007general,davies1974markovian} we transform to the interaction picture relative to the free Hamiltonian $\hat{H}\b t+\hat{H}_B$. We assume the conditions are such that the inertial approximation is valid. In this regime the dipole  operator in the interaction representation can be decomposed in terms of the time-independent eigenoperators $\{ \hat F \}$, Eq. (4) and (6) in the main text, as
\begin{equation}
 \tilde{D}\b t = \sum_n a_n \hat{F}_n e^{-i \Lambda_n\b t}~~,
 \label{eq:D inter}
 \end{equation}
 where $\Lambda_n\b t$ is given by
  \begin{equation}
     \Lambda_n\b t\equiv\int_{\theta\b 0}^{\theta\b t}{d\theta'\sb{\lam_{n} -i\b{\v {G}_{n},\nabla_{\v{\chi}} \v F_{n}}\cdot \f{d \v \chi}{d\theta'}}}~~.
     \label{eq:Lambda}
 \end{equation}

 Here, $\hat F_n \equiv \hat{F}_n\b 0$, $a_n=\text{tr}\b{\v{D}\b 0 \hat{F}_{n}^{\dagger}}$ and a up-script tilde designates operators in the interaction picture. 
 Utilizing Eq. \eqref{eq:D inter},  the composite Hamiltonian in the interaction picture can be written as
\begin{multline}
\tilde{H}_{tot}\b t=\tilde{H}_I \b t\\= i\sum_{\v{k},\lambda,n}\sqrt{\f{ \hbar \omega_k}{2 \eps_0 V}}\v{e}_k\brac{\v{k}}a_n \hat{F}_n e^{-i \Lambda_n \b t}\\\times  \b{b_k\brac{\v{k}}e^{-i\omega_k t}+b_k^{\dagger}\brac{\v{k}}e^{i\omega_k t}}~~.
\label{eq.1}
\end{multline}
We proceed by assuming the Born Markov approximation to obtain the quantum Markovian Master equation 
\begin{equation}
\f{d}{dt}\tilde{\rho}_S\b t=-\f{1}{\hbar^2}\int_0^{\infty}ds\,\text{tr}_B \sb{\tilde{H}_I\b t,\sb{\tilde{H}_I\b{t-s},\tilde{\rho}_S\b t\otimes \tilde{\rho}_B}}~,
\label{eq.2}
\end{equation}
where $\hat{\rho}_B$ is the density operator of the bath.
Assuming the bath correlation functions decay fast relative to the external driving we approximate $\Lambda_k\b{t-s}$ as
\begin{equation}
 \Lambda_k\b{t-s} \approx \Lambda_k\b t- \alpha_k\b t s~~,
 \label{eq.3}
 \end{equation} 
 where $\alpha_k \b t \equiv \Omega_k\b t\lam_k\b t-i\b{\v {G}_{k}\b t|\nabla_{\v{\chi}} \v F_{k}\b t\cdot \f{d \v \chi}{dt}}$.
 This approximation is justified, as the bath correlation functions decay in a typical timescale which is much smaller than the timescale of the change in the system parameters, namely, the function $\Lambda_k \b t$. Thus, the contribution to the integral in Eq. \eqref{eq.2} vanishes when the approximation Eq. \eqref{eq.3} is not satisfied, see \cite{dann2018time} for further details. \\
Gathering  equations \eqref{eq.1}, \eqref{eq.2} and \eqref{eq.3} leads to
\begin{multline}
 \f{d}{dt} \tilde{\rho}_S \b t= \sum_{i,j} e^{-i\sb{\Lambda_i\b t+\Lambda_j \b t}}\Gamma_{ij}\b{\alpha_j\b t}\\ \times \b{\hat F_j \tilde{\rho}_S\b t \hat F_i -\f{1}{2}\{\hat F_i \hat F_j \tilde{\rho}_S\b t\}}~~,
\label{eq.4}
\end{multline}
with the spectral correlation tensor given by
\begin{equation}
\Gamma_{ij}\b{\alpha_j \b t}=\f{a_i a_j}{\hbar^2}\int_0^{\infty}ds\,e^{i \lam_j\alpha \b t s}\mean{E_i\b t E_j\b{t-s}}~~.
\end{equation}
We assume $\Lambda_i\b t+\Lambda_j\b t\gg1$ for $\Lambda_i\b t\neq-\Lambda_j\b t$, and by performing the secular approximation terminate terms in Eq. \eqref{eq.4} which oscillate rapidly.
Furthermore, by following a similar derivation as presented in Ref. \cite{breuer2002theory} Part. II Sec. 3.4.1,  the spectral correlation tensor $\Gamma_{ij}$ can be calculated and written as a sum of two terms  $\Gamma_j\b{\alpha}\equiv\Gamma_{ij} =\delta_{ij}\b{\f{1}{2}\gamma_j \b{\alpha}+i S_j \b{\alpha}}$, with
\begin{equation}
\gamma_j\b{\alpha } = \f{ { \alpha}^3|a_j\v{d}|^2}{12  \pi^2 \hbar \eps_0  c^3}\b{1+N\b{\alpha}}~~,
\end{equation}
and 
\begin{equation}
S_j\b{\alpha} = \f{|a_j\v{d}|^2}{6\pi^2 \hbar \eps_0 c^3} {\cal{P}} \int_0^\infty d \omega_k \omega_k^3 \sb{\f{1+N\b{\omega_k}}{\alpha-\omega_k}+\f{N\b{\omega_k}}{\alpha+\omega_k}}~~.
\end{equation}
Here, $c$ is the speed of light, ${\cal{P}}$ designates the Cauchy principle part and $N\b{\alpha}$ is the occupation number of the Bose-Einstein distribution at frequency $\alpha$. 

The final form of the NAME in the interaction picture can be written as
\begin{multline}
\f{d}{dt} \tilde{\rho}_S \b t= -\f{i}{\hbar}\sb{\tilde{H}_{LS}\b t,\tilde{\rho}_S\b t}\\+ \sum_{j} \gamma_j\b{ \alpha_j\b t} \b{\hat F_j \tilde{\rho}_S\b t \hat F_j^{\dagger} -\f{1}{2}\{\hat F_j^{\dagger} \hat F_j, \tilde{\rho}_S\b t\}}~~,
\label{eq.5}
\end{multline}
where $\tilde{H}_{LS}$ is the Lamb shift correction term in the interaction representation
\begin{equation}
\tilde H_{LS}\b t = \sum_j \hbar \alpha_j \b t\hat F_j^\dagger \hat F_j~~.
\end{equation}
Equation \eqref{eq.5} is of the GKLS form guaranteeing a complete positive trace preserving the dynamical map \cite{lindblad1976generators,gorini1976completely,alicki2007general}.

%\trr{Notice that the decay rates, $\Gamma_j$ in Eq. \eqref{eq.5} are a dependent on the frequencies $\alpha_j \b t$. In turn, these are connected to the dynamical and geometric phases through Eq. \eqref{eq.3}. This connection leads to the conclusion that the dynamical and geometric phases influence the energy and information transfer with the bath.}

\section{Eigenoperators and eigenvalues and invariants}
\label{ap:eig}
\paragraph{Parametric harmonic oscillator}
The matrix $\cal{B}$, Eq. (18) in the main text, can be decomposed to two block matrices, the eigenvectors in the $\{ \hat H,\hat L,\hat C\}$ basis of the upper 3 by 3 matrix are $\v F_{1}=\f{1}{\sqrt{4+\mu^2}}\{2,0,\mu\}^{T}$,   $\v F_{2}=\f{1}{\sqrt 8}\{\mu,i\kappa,2\}^{T}$ and   $\v F_3 =\f{1}{\sqrt 8}\{\mu,-i\kappa,2\}^{T}$,
corresponding to the eigenvalues $\lam_1 =0$ ,$\lam_2 =\kappa$  and $\lam_3= -\kappa$. Each eigenvector $\v F_k$ corresponds to the eigenoperator $\hat F_k$, which is obtained by summing over the  product of the coefficients and the basis operators. For $\v{F}_k=\{f_j^1,f_j^2,f_j^3 \}^T$, $\hat{F}_j= f_j^1 \hat{H} + f_j^2 \hat{L}+ f_j^3 \hat{C}$. 
The eigenvectors that correspond to the eigenopertors of the bottom block 2 by 2 matrix are
\begin{equation}
\v F_+=\f{1}{\sqrt{8}}\{2,\mu+i \kappa\}^T  \,\,\,  \v F_-=\f{1}{\sqrt{8}} \{2,\mu-i \kappa\}^T~~,
\end{equation}
with the eigenvalues $\lam_+=\f{\kappa}{2}$ and $\lam_-=-\f{\kappa}{2}$.
The dynamics of the eigenoperators has an additional scale $\f{\omega\b t}{\omega \b 0}$ associated with the diagonal of terms of $\cal B$. For example,
the dynamics of the operator associated with $\v{F}_1$ (vanishing eigenvalue) is $\hat{F}_1^H\b t = \f{\omega \b t}{\omega 0}\hat{F}_1 \b 0$, where the upper-script H designates that the operator is in the Heisenberg picture. Explicitly, the dynamics becomes
\begin{equation}
    \hat{F}_1^H\b t= \f{\omega\b t}{\omega \b 0}\f{1}{\sqrt{4+\mu^2}}\b{2\hat{H}\b 0+\mu\hat{C}\b 0}~~.
    \label{apeq:HO_invariant}
\end{equation}

The geometric phase (Eq. (7) in the main text) is obtained by integrating over the Berry connection: $i \b{\v{G}_{k},\nabla_{\v{\chi}} \v F_{k}}=i \b{\v{G}_{k},\partial_{\mu} \v F_{k}}$. This expression vanishes for the eigenvectors of the harmonic oscillator, therefore, the inertial solution does not contain a geometric phase.

\paragraph{Two-Level-System}
The eigenvectors, that correspond to the eigenoperators, and eigenvalues of the propagator (Eq. (23) in the main text) are $\v F_1=\f{1}{\bar{\kappa}}\{1,0,\bar{\mu}\}^{T}$, $\v F_2=\f{1}{\sqrt2\bar{\kappa}}\{-\bar{\mu},-i\bar{\kappa},1\}^{T}$  and   $\v F_3 =\f{1}{\sqrt2\bar{\kappa}}\{-\bar{\mu},i\bar{\kappa},1\}^{T}$,
with the eigenvalues $\lam_1=0$,   $\lam_2=\bar{\kappa}$,  $\lam_3 =-\bar{\kappa}$, where $\bar{\kappa}=\sqrt{\bar{\mu}^2+1}$. The operators proportionate to the invariant $\v{F}_1$ is
\begin{equation}
    \hat{F}_1\b t= \f{1}{\bar{\kappa}}\b{\bar{H}\b t+\bar{\mu}\bar{C}\b t}~~,
    \label{apeq:TLS_invariant}
\end{equation}
where the invariant is $\hat{G}_1\b t$, given in Eq. \eqref{eq:g_apend}.

In a similar fashion as in the harmonic oscillator model, the Berry connection, associated with the various eigenvectors, vanishes.

\paragraph{Three-Level-System}
The dynamics of the inertial solution $\v v^H\b t$ in the $\{\{\hat{T}\},\hat{I}\}$ basis in Liouville space is generated by $\f{d}{dt}\v v^H= -i{\cal{M}}\b t\v v^H$, where ${\cal M}\b{\chi_1\b t,\chi_2\b t}$ is given by
\begin{equation}
{\cal M}= i\Omega\sb{\begin{array}{cccccccc}
0 & -\chi_1 & 0 & 0 & 0 & -\chi_2 & 0 & 0\\
\chi_1 & 0 & 2 & 0 & 0 & 0 & \chi_2 & 0\\
0 & -2 & 0 & -\chi_2 & 0 & 0 & 0 & 0\\
0 & 0 & \chi_2 & 0 & 0 & 0 & 1 & \sqrt{3}\chi_2\\
0 & 0 & 0 & 0 & 0 & -1 & 0 & 0\\
\chi_2 & 0 & 0 & 0 & 1 & 0 & \chi_1 & 0\\
0 & -\chi_2 & 0 & -1 & 0 & -\chi_1 & 0 & 0\\
0 & 0 & 0 & -\sqrt{3}\chi_2 & 0 & 0 & 0 & 0
\end{array}}~~,
\label{eq:3level_M}
\end{equation}
 with $\chi_1=\Delta/\Omega$ and $\chi_2=\b{\alpha\dot{\beta}-\dot{\alpha}\beta}/{\Omega^3}$ is the adiabatic parameter.

The eigenvectors in Liouville space with vanishing eigenvalues correspond to invariant observables. The three-level-system (described in section 3 C in the main text) has a set of two independent invariants. Any linear combination of these operators is also an invariant of the dynamics.  These invariants form a vector space, spanned by the vectors (given in the ${\hat{T}}$ basis)
\begin{gather}
    \v e_{1}=N_{1}\left\{\sqrt{3},0,-\frac{\sqrt{3} \chi_1 }{2},0,-\sqrt{3} \chi_2 ,0,0,\frac{\chi_1
   }{2}\right\}
   \\ \v e_{2} = N_{2}\Big\{ \chi_1-8\chi_1\chi_2^{2},0,\chi_1^{2}\left(3\chi_2^{2}-\frac{1}{2}\right)\\-3\chi_2^{2}\left(\chi_2^{2}+1\right),
   0,\chi_1\chi_2\left(-2\chi_1^{2}+2\chi_2^{2}-7\right),0,\nonumber\\2\chi_2\left(\chi_1^{2}+3\chi_2^{2}+3\right),-\frac{1}{2}\sqrt{3}\left(2\left(\chi_1^{2}+1\right)\chi_2^{2}+\chi_1^{2}-2\chi_2^{4}+4\right)\Big\} ~,\nonumber
\end{gather}
where $N_1$ and $N_2$ are the normalization factors.
The vector representing the invariant $\hat{C}$ is given by $\v{C}=\b{2\sqrt{\chi_1^{2}+3\chi_2^{2}+3}}^{-1}\b{c_1\v e_1+c_2 \v e_2}$, with 
\begin{gather}
    c_1=\b{-6\left(\chi_1^{2}-1\right)\chi_2^{2}+\chi_1^{2}+6\chi_2^{4}}/{\left|\chi_1\right|}\\
    c_2= -\sqrt{12\chi_1^{4}\chi_2^{2}+3\chi_1^{2}\left(-8\chi_2^{4}+20\chi_2^{2}+1\right)+12\left(\chi_2^{2}+1\right)^{3}}~.
    \nonumber
\end{gather}
The vector corresponding to the invariant $\hat{D}$ is $\v{D}=\b{2\sqrt{\chi_1^{2}+3\chi_2^{2}+3}}^{-1}\b{c_2\v e_1-c_1 \v e_2}$.

\section{Construction of inertial Hamiltonians - detailed derivation}
\label{apsec:generalization}
We begin by presenting a detailed derivation of the wave-function propagator $\hat{U} \b{t,0}$, starting from the inertial solution in Liouville space. The inertial solution  of the two-level system, Sec. 3 B, leads to three othonormal operators which are proportionate to the eigenoperators $\{\hat{F}_1,\hat{F}_2,\hat{F}_3\}$: 
\begin{gather}
    \hat{G}_{1}\b{t}=\f{\sqrt{2}}{\bar{\Omega}\b t\bar{\kappa}}
    \b{\bar{H}\b t+\bar\mu \bar{C}\b t}\nonumber\\
    \hat{G}_{2}=\hat{G}_3^\dagger\b t=\f 1{\bar{\Omega}\b t\bar \kappa}\b{-\bar\mu \bar{H}\b t-i\bar \kappa \bar{L}\b t+\hat{C}\b t}~~,
    \label{eq:g_apend}
\end{gather}

The vec-ing procedure, described in Sec. 5 in the main text, implies that $\hat{G}_2$ and $\hat{G}_3$ are of the form $\ket{\phi_i\b t}\bra{\phi_j\b t}$, where $i,j=1,2$ with $i\neq j$, and $\{\ket{\phi_i\b t}\}$ are the eigenstates of the $\hat{U}\b{t,0}$. This structure allows obtaining the eigenstates be diagonalization of $\hat{G}_2$ or $\hat{G}_3$, substituting the TLS operators $\{\bar{H},\bar{L},\bar{C}\}$, into Eq. \eqref{eq:g_apend} leads to the eigenstates
\begin{gather}
    \ket{\phi_{1}\b t}=\sqrt{\f{\bar \kappa\bar{\Omega}+\omega}{2\bar \Omega\bar \kappa}}\left\{ i\frac{\bar \mu\bar \Omega+i\epsilon}{\bar \kappa\bar \Omega+\omega},1\right\}^T \nonumber\\
    \ket{\phi_{2}\b t}=\sqrt{\f{\bar \kappa\bar \Omega-\omega}{2\bar \kappa\bar \Omega}}\left\{ -\frac{i\left(\bar \mu\bar \Omega+i\epsilon\right)}{\bar \kappa\bar \Omega-\omega},1\right\}^T~~,
\end{gather}
where all the Hamiltonian parameters may be time-dependent. In turn of the eigenstates, the operators are given by $\hat{G}_{2}\b t=\hat{G}_3^\dagger\b t = e^{i\lam\b t}\ket{\phi_{1}\b t}\bra{\phi_{2}\b t}$, where $\lam\b t\f{i\eps\bar \kappa+\bar \mu\omega}{\sqrt{\eps^{2}\bar \kappa^{2}+\bar \mu^{2}\omega^{2}}}$ and $\hat G_{1}=\f 1{\sqrt{2}}\b{\ket{\phi_{2}\b t}\bra{\phi_{2}\b t}-\ket{\phi_{1}\b t}\bra{\phi_{1}\b t}}$.
We can now compare the dynamics of the expectation values 
\begin{multline}
    \mean{\hat{G}_{2}\b t}=e^{i\lam\b t}\text{tr}\b{\ket{\phi_{1}\b t}\bra{\phi_{2}\b t}\hat{U}\hat{\rho}\b 0\hat{U}^{\dagger}} \\ = e^{i\lam\b t}\text{tr}\b{\hat{U}^{\dagger}\ket{\phi_{1}\b t}\bra{\phi_{2}\b t}\hat U\hat{\rho}\b 0} \\=e^{i\lam\b t} \bra{\phi_{1}\b 0}\hat U^{\dagger}\ket{\phi_{1}\b t}\bra{\phi_{2}\b t}\hat{U}\ket{\phi_{2}\b 0}\\ \times \bra{\phi_2\b 0}\hat{\rho}\b 0\ket{\phi_1\b 0},
    \label{eqap:G_derivation}
\end{multline}
where the last equality stems from the fact that $\hat{U}\b t$ is a sum of terms of the form $\ket{\phi_k\b t}\bra{\phi_k\b 0}$ (Eq. (38) in the main text).

We now compare Eq. \eqref{eqap:G_derivation} with the result obtained from the inertial solution in Liouville space:
\begin{multline}
        \mean{\hat{G}_{2}\b t}=\text{tr}\b{\hat{G}^{H}_{2}\b t\hat{\rho}\b 0}=e^{-i\int_0^t\bar{\kappa}\bar\Omega dt'}\mean{\hat{G}_{2}\b 0}
        \\=e^{-i\int_0^t\bar{\kappa}\bar\Omega dt'}e^{i\lam\b 0}\bra{\phi_2\b 0}\hat{\rho}\b 0\ket{\phi_1\b 0}~~.
    \label{eqap:G_derivation2}
\end{multline}
The results of Eqs. \eqref{eqap:G_derivation} and \eqref{eqap:G_derivation2} determine, up to a global phase, the phases of the propagator $\alpha_k$, Eq. (8) in the main text, and the propagator becomes
\begin{equation}
    \hat{U}\b{t,0}=e^{i\Lambda\b t}\ket{\phi_{1}\b t}\bra{\phi_{1}\b 0}+e^{-i\Lambda\b t}\ket{\phi_{2}\b t}\bra{\phi_{2}\b 0}~~.
\end{equation}

\subsection*{Construction of a three-level-system Hamiltonian }
By combining two TLS propagators we obtained an inertial Hamiltonian for a three-level system. We considered two propagators that couple two energy levels with a single common energy state, which is taken to be the second energy state. The new Hamiltonian is given by (Eq. (41) in the main text)
\begin{equation}
    \hat{H}\b t = \hat{H}^{\b{1}}\b{t}+\hat{H}^{\b{2}}\b{t}+\hat{H}_{12}\b t
\end{equation}
where $\hat{H}^{i}=\omega^{\b{i}}\b t \hat{S}^{\b{i}}_z+ \eps^{\b{i}}\b t \hat{S}^{\b{i}}_x$. The correction term is $\hat{H}_{12}\b t = \sb{\hat{U}^{\b{1}}\b{t,0},\hat{H}^{\b{2}}\b{t}}\hat{U}^{\b{1}\dagger}\b{t,0}$, where the commutation relation is given by 
\begin{widetext}
\[
%\begin{equation}
%\begin{equation}
\sb{\hat{U}^{\b{1}},\hat{H}^{\b{2}}}=
\left(\begin{array}{ccc}
0 & -\omega^{\b 2}F\b{f_{+}\b ty_{+},f_{-}\b ty_{-}} & -\eps^{\b 2}F\b{f_{+}\b ty_{+},f_{-}\b ty_{2}}\\
\omega^{\b 2}F\b{f_{+}^{*}\b 0y_{+},f_{-}^{*}\b 0y_{-}} & 0 & \eps^{\b 2}\left(F\b{y_{+},y_{-}}-\f 12\right)\\
\eps^{\b 2}F\b{f_{+}^{*}\b 0y_{+},f_{-}^{*}\b 0y_{-}} & -\eps^{\b 2}\left(F\b{y_{+},y_{-}}-\f 12\right) & 0
\end{array}\right)
\nonumber
%\end{equation}
%\end{equation}
\]
\end{widetext}
with $F\b{x,y}=\f 12\b{xe^{i\Lambda^{\b{1}}}+ye^{-i\Lambda^{\b{1}}}}$ and
\begin{gather}
    f_{\pm}\b t=\f{\eps^{\b{1}}\b t-i\mu^{\b{1}}\Omega^{\b{1}}\b t}{\kappa^{\b{1}}\Omega^{\b{1}}\b t\pm\omega^{\b{1}}\b t}\nonumber\\
    g_{\pm}\b t=\sqrt{\f{\kappa^{\b{1}}\Omega^{\b{1}}\b t\pm\omega^{\b{1}}\b t}{2\kappa\Omega^{\b{1}}\b t}}\nonumber\\
    y_\pm\b t=g_{\pm}\b tg_{\pm}\b 0\\
   \Lambda^{\b{1}}\b t=\int_0^t\bar \kappa\bar\Omega dt'+\f{\lam^{\b 1}\b t-\lam^{\b 1}\b 0}2~~.
\end{gather}

\section{Geometric phase example}
\label{ap:geo example}

The inertial solution of Hamiltonian (Eq. (46) in the main text) includes two non-vaninshing geometric phases. The solution, Eq. (6) in the main text, is obtained by introducing a scaled time $\tau\b t=\int_0^t\,t' f\b{t'}$ and diagonalizing ${\cal{H}}$.  The eigenvectors in Liouville space read
\begin{equation}
    \v F_{0}=\{\sin\theta \cos\varphi,\sin\theta \sin\varphi,\cos\theta\}^T\\
    \nonumber
\end{equation}
\begin{equation}
    \v F_{\pm}=\f 1{\sqrt{2}}\big\{ -\cos\theta \cos\varphi\pm i\sin\varphi,-\cos\theta \sin\varphi\mp i\cos\varphi,\sin\theta\big\}^T~~,
    \label{eqap:eigenvectors}
\end{equation}
where $\theta$ and $\varphi$ are the spherical angles describing the direction of the magnetic field $\v B \b t$. The corresponding eigenvalues are $\lam_0=0$ and $\lam_\pm\b t=\pm |\v{B}\b t| \bar{\Omega}$. The geometric phase $\phi_k$ is obtained by integrating over the Berry connection (Eq. (7) in the main text), 
\begin{equation}
    A_k=  i\b{\v F_{k},\nabla_{{\theta,\varphi}}\v F_{k}}=i\{\b{\v F_{k},\partial_{\theta}\v F_{k}},\b{\v F_{k},\partial_{\varphi}\v F_{k}}\}~.
    \label{eqap:A}
\end{equation}
Substituting Eq. \eqref{eqap:eigenvectors} into Eq. \eqref{eqap:A} leads to the associated geometric phases: $\phi_0=0$ and $\phi_\pm\b{\theta\b t,\varphi \b t} =\pm\cos\b{\theta\b t}\b{\varphi\b t-\varphi\b 0}$.

% \bibliographystyle{ieeetr}

%\bibliography{references.bib}



\end{document}